
\input harvmac.tex
\Title{\vbox{\baselineskip12pt
\hbox{USC-93-025/YCTP-P22-93}}}
{\vbox{\centerline{Representations of the Virasoro algebra}
\vskip2pt\centerline{from lattice models}}}

\centerline{W.M.Koo\foot{On leave from Physics Dept., Yale University, New
Haven CT 06511}$^\dagger$, H.Saleur\foot{Packard Fellow}
$^\spadesuit$}
\vskip2pt
\centerline{$^\dagger$Department of Physics }
\centerline{University of southern California}
\centerline{Los Angeles CA 90089}
\vskip2pt
\centerline{$^\spadesuit$Department of Physics and Department of Mathematics}
\centerline{University of southern California}
\centerline{Los Angeles CA 90089}
\vskip.3in

We investigate in details how the
Virasoro algebra appears in the scaling limit  of the simplest
lattice models of XXZ or RSOS type. Our approach is straightforward
but to our knowledge had never been tried so far. We simply formulate
a conjecture for the lattice stress-energy tensor motivated by the exact
 derivation of lattice global Ward identities. We then check that
 the proper algebraic relations are obeyed in the scaling limit.
The latter is under reasonable control  thanks
to the Bethe-ansatz solution. The results, which are mostly numerical
for technical reasons,
 are remarkably precise. They are also corroborated by exact
 pieces of information from various sources,
in particular Temperley-Lieb algebra representation theory.
Most  features of the Virasoro algebra (like central term, null vectors, metric
properties...)
 can thus be observed using the
lattice models. This seems of general interest for  lattice field theory, and
also more specifically for  finding relations between conformal invariance  and
lattice integrability,  since
 basis for the irreducible representations of the Virasoro algebra  should now
follow (at least in principle) from
Bethe-ansatz computations.

\newsec{Introduction}

 In the last
 years there has been much progress  in the study of two dimensional
integrable lattice models  and integrable field theories. Deep analogies
 have been discovered  between the two subjects.

It is certainly expected that some relations exist between integrable
lattice models
and integrable field theories  because of their many  common mathematical
ingredients. Such relations
are  often straightforward; for instance R matrices appear both in the
construction of Boltzmann weights \ref\RevI{M.Jimbo, ``Yang-Baxter
 equation in integrable systems'', Advanced Series in Mathematical Physics Vol.
11, World Scientific, and references therein.} and the study
 of braiding properties of
conformal fields \ref\RevII{ A.Tsuchiya, Y.Kanie,
Lett. Math. Phys. 13 (1987) 303; G.Felder, J.Fr\"ohlich, G.Keller,
 Comm. Math. Phys. 124 (1989) 647;
 G.Moore, N.Seiberg, Phys. Lett. B212 (1988) 451;
 L.Alvarez Gaum\'e, C.Gomez and  G.Sierra, Nucl. Phys. B319 (1989) 155.}
 simply because of
 the ubiquity of the Yang-Baxter equation.
There are however deeper relations that are not yet understood.
For instance in \ref\EJMO{E.Date, M.Jimbo, T.Miwa, M.Okado, Phys. Rev.
B35 (1987) 2105.}
 it was shown  that local height probabilities (or order parameters)
in restricted solid on solid (RSOS) integrable
lattice models can generally be  expressed
 in terms of branching functions of conformal field theories.
A consequence of this observation is that the logarithm of
the corner transfer matrix
\ref\B{R.J.Baxter, ``Exactly solved
 models in statistical mechanics'', Academic Press, London (1982).} provides a
 ``lattice Virasoro generator $L_0$''. More precisely  observe first
that the  natural
 configuration space
 of RSOS models is a   space of paths on restricted weight lattices
 \ref\RevIII{M.Jimbo, T.Miwa, M.Okado, Lett. Math. Phys. 14 (1987) 123.}.
For any representation  of the Virasoro  algebra (we restrict to this
 algebra for simplicity) there exists a choice of boundary
 conditions for the RSOS paths such that, restricted to these paths,
the trace of the exponential of the  corner transfer matrix coincides
with the Virasoro  character. This may be some
sort of coincidence due to the common presence of elliptic functions in the
parametrization of off-critical integrable Boltzmann weights and in
the branching functions.
However the final formula for $L_0$ and the path structure of the
potential Virasoro representations are so
 beautiful that one hopes there exists in fact a complete action of the
 Virasoro algebra on the space of paths. This would be physically
 intriguing because
 the Virasoro algebra is traditionnally associated
with conformal invariance, a property of critical continuum theories,
while we deal
here with discrete, off-critical models. This would also
be mathematically useful as the RSOS paths probably would provide
very simple basis for the irreducible Virasoro modules and maybe play a
role similar
to the crystal basis \ref\Kash{M.Kashiwara, Duke Math. J. 69 (1993) 455.}. The
action
 of the Virasoro algebra
on RSOS paths  has  only  been exhibited
in the  Ising model using the underlying fermions \ref\IT{H.Itoyama,
H.B.Thacker, Phys. Rev.
 Lett. 58 (1986) 1395.}\ref\C{J.Cardy, unpublished}\ref\S{H.Saleur,
in ``Knots, topology and
quantum field theories", ed. L.Lusanna, World Scientific (1989).} so far,
although
there has been  much  progress
 recently in understanding the spectrum of
 corner transfer matrices
\ref\ctmi{B.Davies, O.Foda, M.Jimbo, T.Miwa,
A.Nakayashiki, Comm. Math. Phys. 151 (1993) 89.}
\ref\ctmii{H.Frahm, H.B.Thacker, J.Phys. A24 (1991) 5587.}.
The possible  action of a current algebra on paths has been studied in
\ref\Date{
E.Date, M.Jimbo, A.Kuniba, T.Miwa and M.Okado, in Proceedings of the Taniguchi
conference ``Integrable models in quantum field theory and statistical
mechanics'',
Advanced Studies in Pure Mathematics.}.

Close relations between integrable lattice models and conformal field theories
 have also been observed is the
spectrum of row to row transfer matrices in critical models. In
\ref\PS{V.Pasquier, H.Saleur,
 Nucl. Phys. B330 (1990) 523.} (see also \ref\AGR{F.Alcaraz, U.Grimm,
 V.Rittenberg, Nucl. Phys. B316 (1989) 735.}) it was found for instance that
 degeneracies in the free field representations of the Virasoro
 algebra could be observed as well on finite lattice systems of XXZ type.
Qualitatively, the
 lattice quantum group plays  a role similar to   the
screening operators algebra \ref\DF{Vl.S.Dotsenko, V.A.Fateev,
Nucl. Phys. B240 (1984) 312.} and its commutant, the Temperley-Lieb algebra,
 a role similar  to the Virasoro algebra. So far, only Virasoro characters
have been  reproduced, by taking a `scaling limit'' (to be discussed
in details later)  and using
for  ``lattice $L_0$'' the logarithm of the row to row transfer matrix.
The situation is thus rather similar to the one we discussed previously for
off-critical models and corner transfer matrices,
 and the most intriguing question is again  to find some  expressions
 for the other
Virasoro generators.  A first  difference is that the results
obtained from the row to row transfer matrix are not as nice as the ones
ones from the
corner transfer matrix.
Another  difference
  is that we  deal now with a  critical system, where appearance of
the Virasoro algebra in the scaling  limit is expected. The search
for the Virasoro algebra in the lattice model \ref\Tha{H.B.Thacker, Physica 18D
(1986) 348.} gains thus important physical
motivations. Some natural questions to ask  are:
is there some nice lattice deformation of the Virasoro algebra?
 is there a  reformulation of
lattice integrability as some  sort of lattice conformal invariance?
 are there
lattice Ward identities?

In this work  we investigate in details how the
 Virasoro algebra appears in the scaling limit of the simplest
 lattice models of XXZ or RSOS type, and give a partial answer
to the third question. For a different approach related to the same problem see
\ref\McCoyetal{R. Kedem, T.R. Klassen, B.M.
McCoy, E. Melzer, Phys.Lett. B307 (1993) 68.}.
The paper is organized as follows. In section 2 we discuss
 the concept of lattice stress energy tensor.
By generalizing the ideas of Kadanoff and Ceva we study the behaviour
 of the square lattice Q-state Potts model under stretching and straining
and  we write global lattice Ward identities.
This allows us to make a reasonable guess for the local stress energy tensor.
We reformulate this  algebraically using the Temperley-Lieb algebra formalism.

The strategy is then to simply  compute  commutators of our
lattice expressions and study their behaviour for large systems.

 Section 3 contains  first the detailed example of the Ising model where
as usual things work out nicely.
 We then discuss in some generality what kind of convergence can
 be expected for the lattice quantities, and we define precisely the
 scaling limit that
 should be taken. Sections 4 and 5 contain a detailed study of
the XXZ chain. In that case the lattice model behaves like the Coulomb gas with
a charge at infinity and provides non-unitary representations of the Virasoro
algebra. We check numerically that our lattice expressions disentangle the Left
and Right sectors and reproduce the necessary commutation relations,
in particular
the central term and the null vectors structure. Section 6 contains
a similar study for
the periodic RSOS models. In that case the Temperley-Lieb
algebra representation is unitary and so is the Virasoro
representation obtained. Section 7 contains finally a
discussion of the case of fixed boundary conditions for the RSOS models.
Technically this is the most difficult case because convergence of
 numerical data is not as good as for periodic boundary conditions.
It is however far better from a conceptual point of view since in that
 case there is a single Virasoro algebra and by appropriate choice
of boundary conditions a single irreducible representation can be
 selected. Section 8 contains final comments, in particular
about the possibility of making the scaling limit simpler  by
considering time dependent correlation functions.

\newsec{Lattice stress energy tensor}

We now  discuss the concept of lattice stress
energy tensor for critical integrable lattice models based on
the Temperley Lieb algebra. Our arguments are a simple extension of
 the pioneering work of Kadanoff and Ceva \ref\KC{L.P.Kadanoff, H.Ceva,
Phys. Rev. B11 (1971) 3918.}.

\subsec{$Q$ state Potts model: generalities}

 We consider the Q-state Potts model \B\ on a  square lattice whose
vertices are labelled by a pair of integers $(j,k)$. To each vertex is
thus associated a spin variable $\sigma_{jk}$
that takes values $1,\ldots,Q$. The exponential statistical weight
has the form
\eqn\weig{A=\sum_{jk}K_x(j+1/2,k)\delta(\sigma_{jk},\sigma_{j+1,k})+
K_y(j,k+1/2)\delta(\sigma_{jk},\sigma_{j,k+1}),}
where the couplings are in general edge dependent  and the natural
labelling of edges is illustrated in figure 1a. The usual partition
function is
\eqn\parti{Z(\{K\})=\sum_{\sigma_{jk}}e^A,}
where the notation indicates dependence on all the
couplings and  the sum is taken over all possible values for each spin.
We introduce a more properly  normalized partition function as
\eqn\normpar{Y(\{K\})=\prod_{\hbox{edges}}\left({\sqrt{Q}\over
e^K-1}\right)^{1/2}Z,}
where the  notation means that a product over all edges $(j+1/2,k)$ and
$(j,k+1/2)$
is taken and for each, the proper coupling constant put in denominator. By
standard
high temperature expansion techniques the Q-state Potts model can be defined
for any real
$Q$ \B\ .

An important property is the duality symmetry. Define the dual lattice
as the square lattice whose vertex $(j,k)$ stands in the middle of the face
of the original lattice with south west corner $(j,k)$ (see
figure 1b). Define the mapping $K
\rightarrow K^\star(K)$ by
\eqn\map{{e^{K^\star(K)}-1\over\sqrt{Q}}={\sqrt{Q}\over e^K-1}.}
Define accordingly couplings on the dual lattice
\eqn\dcou{K_x^\star(j+1/2,k)=K^\star(K_y(j+1,k+1/2)),}
and
\eqn\decoui{
K_y^\star(j,k+1/2)=K^\star(K_x(j+1/2,k+1)).}
The dual of the original Potts model is the model with couplings \dcou\ ,
\decoui\ defined on the dual lattice. As follows from the standard analysis \B\
the partition functions
 for  the original model  and its dual are simply related: for large systems
where
 boundary effects
can be neglected  one has
\eqn\dual{Y(\{K\})=Y(\{K^\star(K)\}).}

We shall call {\bf homogeneous} the case  where couplings  depend on
 $x,y$ but
not on the position. It is  convenient to introduce the variables
\eqn\newvar{\epsilon_x\equiv{e^{K_x}-1\over\sqrt{Q}},\
\epsilon_y\equiv{e^{K_y}-1\over\sqrt{Q}}.}
One has then
\eqn\newpar{Y(\{K\})\equiv Y(\epsilon_x,\epsilon_y)=Y(\epsilon_y,\epsilon_x)=
Y\left({1\over \epsilon_y},{1\over\epsilon_x}\right),}
where the last two equalities hold for a big system and follow respectively
from $x\leftrightarrow y$ symmetry and duality.

In the homogeneous case the model
 is {\bf critical} when \B\
\eqn\crit{K_x=K^\star(K_y),\ K_y=K^\star(K_x).}
We are interested only in second order phase transitions
 which occur for $0\leq Q\leq 4$. It is  convenient to parametrize then
\eqn\parai{\sqrt{Q}=2\cos\gamma,\gamma\in [0,\pi/2],}
and
\eqn\paraii{\epsilon_y={1\over\epsilon_x}\equiv{\sin u\over
\sin(\gamma-u)},}
where $u$ is called  spectral parameter. We shall restrict to
 the ferromagnetic regime $K_x,K_y>0$
for which   $u\in[0,\gamma]$.

\subsec{Lattice $T_{xx}$}

When the model is at  a second order
phase transition point, all correlation functions  at large distance
depend on a single variable
with elliptic like symmetry
\eqn\ell{{\cal R}^2={j^2\over {\cal S}^2}+{\cal S}^2k^2.}
The parameter ${\cal S}$ can be determined in various ways. To start,
we simply borrow the result
of \ref\KP{D.Kim, P.A.Pearce, J.Phys.
A20 (1987) L451.}
\eqn\ss{{\cal S}^2=\hbox{cotan}\left({\pi u\over 2\gamma}\right).}
As a result, a derivative with respect to ${\cal S}$ is equivalent to
 {\bf straining } the system since for functions that depend on \ell\ one
has
\eqn\strai{{\cal S}{\partial\over \partial{\cal S}}
=-j{\partial\over\partial j}
+k{\partial\over\partial k}.}
Refer now to the relative roles of
temperature and energy for analogy: the integral over the system
of the energy density is an ``operator'' which acts (when inserted in
correlation
functions) like a derivative with respect
to temperature. Look here for an operator which acts as
\eqn\oper{<\left({\cal T}_1-<{\cal T}_1>\right)
\sigma_{j_1k_1}\sigma_{j_2k_2}>=\left(
-j_1{\partial\over\partial j_1}+
k_1{\partial\over\partial k_1}-j_2{\partial\over
\partial j_2}+k_2{\partial\over\partial k_2}\right)<\sigma_{j_1k_1}
\sigma_{j_2k_2}>,}
and similarly for any  local quantity other than the spin $\sigma$. In this
section
 $<.>$ indicates  thermal average. Relation \oper is expected to hold in the
limit
where $j,k>>1$ so these variables can be treated as continuous.
{}From \strai\ we can perform the right hand side operation
in \oper\ simply by taking instead derivatives with respect to ${\cal S}$. From
\ss\ this is equivalent to taking derivatives with respect to $u$, and from
\newvar\ and \paraii\ this is also equivalent to taking  derivatives
 with respect to $K_x,K_y$.
But such derivatives
can be taken by acting with an operator that depends on the spins only.
Explicitely things are as follows: one finds first
\eqn\eqi{{\cal S}{\partial\over \partial {\cal S}}={4\gamma\over\pi}
{1\over {\cal S}^2+{\cal S}^{-2}}{\sin\gamma\over\sin u\sin(\gamma-u)}\left[
{e^{K_x}-1\over e^{K_x}}{\partial\over\partial K_x}-{e^{K_y}-1\over e^{K_y}}
{\partial\over \partial K_y}\right].}
Now to take derivatives with respect to $K_x,K_y$ one just has to insert
$\delta$ functions of the appropriate spin interaction: for instance
\eqn\exemi{{\partial\over\partial K_x(j+1/2,k)}<\sigma_{j_1k_1}\sigma_{j_2k_2}>
=\left[<\delta(\sigma_{jk},\sigma_{j+1,k})-<\delta(\sigma_{jk},\sigma_{j+1,k})>
\right]\sigma_{j_1k_1}\sigma_{j_2k_2}>,}
and similarly for any local quantity.  Thus introducing
$$
t_1(j+1/2,k+1/2)\equiv{4\gamma\over\pi}
{1\over {\cal S}^2+{\cal S}^{-2}}{\sin\gamma\over\sin u\sin(\gamma-u)}
$$
\eqn\tttt{
\left[{e^{K_x(j+1/2,k)}-1\over e^{K_x(j+1/2,k)}}\delta(\sigma_{jk},
\sigma_{j+1,k})-{e^{K_y(j,k+1/2)}-1\over
e^{K_y(j,k+1/2)}}\delta(\sigma_{jk},
\sigma_{j,k+1})\right],}
(defined up to an additive constant) one has
\eqn\tttti{{\cal T}_1=\sum_{jk}t_1(j+1/2,k+1/2).}
Obviously ${\cal T}_1$ changes sign in $x\leftrightarrow y$. On the other hand
it is invariant in duality transformations since
$$
{dK^\star\over dK}=-{e^{K^\star}-1\over e^{K^\star}}{e^K\over e^K-1}.
$$

We now want to show that equation  \tttt\ , which can easily be guessed using
the
expected behaviour of ${\cal T}_1$ in $x\leftrightarrow y$
and duality transformations, leads to the result \ss. Observe to start
that the homogeneous lattice model with coupling constants $K_x,K_y$
can be considered at large distance as discretization of an isotropic
 continuum
by rectangles of unit area, whose shape can be characterized by an anisotropy
angle
$\theta$ as shown on figure 2. Clearly one has
\eqn\appi{{\cal S}^2=\tan\left({\theta\over 2}\right).}
The problem is to determine the dependence $\theta(K_x,K_y)$ on the self dual
line \crit\ . To do so, recall from the  definition of the stress energy
tensor in a continuum theory (we use the normalization which is standard
in conformal field theory \ref\BPZ{A.Belavin, A.Polyakov, A.B.Zamolodchikov,
Nucl. Phys. B281 (1984) 333.}) that
\eqn\appii{\delta\ln Z={1\over 2\pi}
\int\left<T_{xx}{\partial\over\partial x}\delta x+T_{yy}
{\partial\over\partial y}\delta y\right>dxdy,}
for a deformation $x\rightarrow x+\delta x$ and $y\rightarrow y+\delta y$.
Considering the  particular deformation $\theta\rightarrow \theta+\delta\theta$
one finds
\eqn\appiii{{\partial\over\partial x}\delta x={1\over 2}
{\cos\theta/2\over\sin\theta/2}\delta\theta,\
{\partial\over\partial y}\delta y=-{1\over 2}
{\sin\theta/2\over\cos\theta/2}\delta\theta.}
Hence, using also $T_{xx}+T_{yy}=0$,
\eqn\appiv{{\delta\ln Z\over\delta\theta}=
{1\over 2\pi\sin\theta}\int <T_{xx}>dxdy.}
For the moment simply observe that the variation of $\ln Z$ is proportional to
the integral of $<T_{xx}>$ which is itself proportional to ${\cal T}_1$ (this
follows from the
 definition
(eg \oper\ ; see also the subsequent discussion of lattive versus conformal
Ward identities). On the other hand, since $\theta$ can be expressed in terms
of $K_x,K_y$
on the self dual line, and assuming this expression is invertible,
one has
\eqn\appv{{\delta\ln Z\over\delta\theta}
=\left(\sum_{jk}a_x{\partial\over\partial K_x}+
a_y{\partial\over\partial K_y}\right)\ln Z,}
where
\eqn\appvi{a_x\equiv{\partial K_x\over\partial\theta},\ a_y\equiv{\partial K_y
\over
\partial\theta}.}
Now comparing \appv\ , \appiv\ and \tttt\ we see that the dependence
$\theta(K_x,K_y)$
must be such that
\eqn\appvii{{a_x\over a_y}=-{e^{K_x}-1\over e^{K_x}}{e^{K_y}\over e^{K_y}-1}.}
It is easy to show that this implies, from parametrization \paraii\ that
$\theta$ is proportional to $\gamma-u$. To determine the proportionality
coefficient simply observe from \paraii\ that the singular values
$\theta=0,\theta=\pi$
 must correspond to $u=0,u=\gamma$. Hence
\eqn\paraviii{\theta={\pi (\gamma-u)\over\gamma}.}

\subsec{Comparison of lattice and  conformal  Ward identities}

We now compare the lattice Ward identity \oper\ with local Ward identities
in conformal field theory. Recall the definitions
\eqn\cftti{T(z)={1\over 4}\left(T_{xx}-T_{yy}-2iT_{xy}\right),}
and
\eqn\cfttii{\bar{T}(\bar{z})={1\over 4}\left(T_{xx}-T_{yy}+2iT_{xy}\right),}
and the local Ward identity
$$
\left<T_{xx}(z,\bar{z})\phi_1(z_1,\bar{z}_1)\ldots,
\phi_N(z_N,\bar{z}_N)\right>=
$$
\eqn\cfttiii{\left(\sum_{i=1}^{i=N}{\Delta_i\over
(z-z_i)^2}+{\bar{\Delta}_i\over
(\bar{z}-\bar{z}_i)^2}
+{1\over z-z_i}{\partial\over\partial z_i}
+{1\over\bar{z}-\bar{z}_i}{\partial\over\partial\bar{z}_i}\right)
\left<\phi_1(z_1,\bar{z}_1)\ldots,
\phi_N(z_N,\bar{z}_N)\right>.}
Now integrate \cfttiii\ over the whole complex plane. Using
$$
\bar{\partial}{1\over z-z_i}=\pi\delta^{(2)}(z-z_i),
$$
one can -formally- establish that
\eqn\intei{\int{dxdy\over z-z_i}=-\pi\bar{z}_i,}
while by derivation
\eqn\inteii{\int{dxdy\over (z-z_i)^2}=0.}
Now use
$$
\bar{z}{\partial\over\partial z}+z{\partial\over\partial\bar{z}}=
x{\partial\over\partial x}
-y{\partial\over\partial y},
$$
to get, integrating \cfttiii\
\eqn\mri{\left<\int  T_{xx}\phi_1\ldots\phi_Ndxdy\right>=
\pi\sum_{i=1}^{i=N}\left(-x_i{\partial\over\partial x_i}
+y_i{\partial\over\partial y_i}\right)\left<\phi_1\ldots\phi_N\right>.}
We could also have obtained this relation directly from the classical
definition of the stress-energy tensor. This looks like  equation \oper\ with
the correspondence
\eqn\corri{{\cal T}_1-<{\cal T}_1>\rightarrow \int
{T_{xx}-T_{yy}\over2\pi}dxdy.}
The only condition for the lattice Ward identity to hold is $j,k>>1$.

So far, only ${\cal T}_1$, a global quantity, has been identified.
We do not know any convincing way
to establish a priori that the density $t_1$ \tttt\ is indeed what will
correspond to
the local quantity $T_{xx}$. We shall assume it is true for the moment, that is
\eqn\resi{t_1(j+1/2,k+1/2)-<t_1>\mapsto{1\over 2\pi}(T_{xx}-T_{yy}).}
Most of the paper
will be devoted to checks of that hypothesis. The precise meaning of the
correspondence indicated by $\mapsto$
 is not as simple as
for the global result \corri. This leads
 to the proper definition of the scaling
limit to be discussed in the next  section. The same remarks applies in all the
rest of this section for expressions involving the notation $\mapsto$.

\subsec{Temperley-Lieb algebra and operator formulation}

To check that $t_1$ corresponds to $T_{xx}$ a first possibility would
 be to investigate
directly local Ward identities for the lattice model. This however requires
the knowledge of many multipoint lattice correlation functions and is therefore
very difficult, analytically or numerically. It is much easier to study the
algebraic relations satisfied by the modes of $t_1$ and compare them with
the Virasoro commutation relations. To do so however we need to pass to a
hamiltonian
description of the lattice model.

Consider the row to row transfer matrix for the Potts model, with time
in the $x$ direction. The transfer matrix is
a true operator acting on configurations of vertical  spins, which we call
 (discarding now the horizontal index) $\sigma_j$, $j=1,\ldots,L$ where $L$
is the size of the system. We choose periodic boundary conditions in space
direction.
 Introduce the operators
\eqn\tli{X_{2j}=1+\epsilon_y e_{2j},\ X_{2j-1}=\sqrt{Q}\left(
\epsilon_x+e_{2j-1}\right),}
where $e_i$ are defined by their matrix elements
\eqn\tlii{\left(e_{2j}\right)_{\sigma,\sigma'}=
\sqrt{Q}\prod_k\delta(\sigma_k,\sigma'_k)\delta(\sigma_j,\sigma_{j+1}),}
(with $\sigma_{L+1}\equiv\sigma_1$) and
\eqn\tliii{\left(e_{2j-1}\right)_{\sigma,\sigma'}=
{1\over\sqrt{Q}}\prod_{k\neq j}\delta(\sigma_k,\sigma'_k).}
They satisfy the relations
\eqn\tldef{e_j^2=\sqrt{Q}e_j,\ e_{j}e_{j\pm 1}e_j=e_j,\ [e_j,e_k]=0,\ |j-k|\geq
2,}
where $e_{2L+1}\equiv e_1$. The transfer matrix itself reads
\eqn\transf{\hat{\tau}=\prod_{j=1}^{j=L} X_{2j}\prod_{j=1}^{j=L} X_{2j-1}.}
In this formalism, $t_1$ takes a rather complicated form in general.
It simplifies however in the hamiltonian limit where   $K_y\rightarrow 0$
and $K_x\rightarrow \infty$. This corresponds to $u\rightarrow 0$. In this
limit $\epsilon_y\approx{u\over\sin\gamma}$.  It is then convenient to proceed
as
follows. Call $\Gamma$ the prefactor in \tttt\ and \eqi\ . Then
$$
\left<t_1(j_1+1/2,k_1+1/2)t_1(j_2+1/2,k_2+1/2)\right>=\Gamma^2
$$
\eqn\tcor{\left[{e^{K_x}-1\over e^{K_x}}{\partial\over \partial K_x}-
{e^{K_y}-1\over e^{K_y}}{\partial\over \partial K_y}\right]\times
\left[{e^{K_x}-1\over e^{K_x}}{\partial\over \partial K_x}-
{e^{K_y}-1\over e^{K_y}}{\partial\over \partial K_y}\right]\ln Z,}
where the labels of the couplings $K$ are implicit. For a general heterogeneous
model the partition function on a rectangle of size $L,T$ with periodic
boundary conditions in time direction as well reads
$$
Z=\hbox{tr}\left(\hat{\tau}\times\ldots\times\hat{\tau}\right),
$$
where the product is over $T$ terms and all the couplings dependence
is implicit. Compute \tcor\ and identify the result with the correlation
function of two
operators introduced at the appropriate places. Using the relations \tldef\
one finds
\eqn\tliv{\hat{t}_1\approx-\Gamma\epsilon_y(e_{2j}+e_{2j-1})+\hbox{constant}.}
To compare with the operator $\hat{T}_{xx}$
a final rescaling  has to be performed since the units of length in $x$ and $y$
direction are different:
\eqn\neweq{\hbox{cotan}\left({\theta\over
2}\right)\left(\hat{t}_1-<\hat{t}_1>\right)\mapsto {1\over
2\pi}\left(\hat{T}_{xx}-\hat{T}_{yy}\right).}
{}From the expression of $\Gamma$
and \resi\ one finds the central result
\eqn\resii{-{2\gamma\over\pi\sin\gamma}(e_{2j}+e_{2j-1}-2e_\infty)
\mapsto{1\over 2\pi}\hat{T}_{xx},}
where $e_\infty$ is the mean value of the $e$ operators in the ground state for
$L\rightarrow\infty$. This value is non universal. Let us emphasize that  in
the conventions used so far,
the system has length $L$. We will change our conventions to $2L$ later.

\subsec{Lattice $T_{xy}$}

So far we have studied only one component of the stress energy tensor. The
other
one should satisfy
\eqn\opertwo{<\left({\cal T}_2-<{\cal T}_2>\right)
\sigma_{j_1k_1}\sigma_{j_2k_2}>=\left(
j_1{\partial\over\partial k_1}+
k_1{\partial\over\partial j_1}+j_2{\partial\over
\partial k_2}+k_2{\partial\over\partial j_2}\right)<\sigma_{j_1k_1}
\sigma_{j_2k_2}>,}
and be even under duality and $x\leftrightarrow y$ transformations. A plausible
expression that satisifes these constraints is
$$
t_2(j+1/2,k)\propto\left[{e^{K_x(j+1/2,k)}-1\over e^{K_x(j+1/2,k)}}{
e^{K_y(j+1,k+1/2)}-1\over e^{K_y(j+1,k+1/2)}}
{\partial\over\partial K_x(j+1/2,k)}{\partial
\over\partial K_y(j+1,k+1/2)}\right.
$$
\eqn\tttttwo{\left.-{e^{K_x(j+1/2,k)}-1\over e^{K_x(j+1/2,k)}}{
e^{K_y(j+1,k-1/2)}-1\over e^{K_y(j+1,k-1/2)}}{\partial\over\partial
K_x(j+1/2,k)}{\partial
\over\partial K_y(j+1,k-1/2)}\right],}
where it is necessary to introduce two derivatives to ensure $x\leftrightarrow
y$
 invariance. Another plausible  expression would be
$$
t_2(j+1/2,k)\propto\left[{e^{K_x(j+1/2,k)}-1\over e^{K_x(j+1/2,k)}}{
e^{K_y(j,k+1/2)}-1\over e^{K_y(j,k+1/2)}}
{\partial\over\partial K_x(j+1/2,k)}{\partial
\over\partial K_y(j,k+1/2)}\right.
$$
\eqn\tttttwo{\left.-{e^{K_x(j+1/2,k)}-1\over e^{K_x(j+1/2,k)}}{
e^{K_y(j,k-1/2)}-1\over e^{K_y(j,k-1/2)}}{\partial\over\partial
K_x(j+1/2,k)}{\partial
\over\partial K_y(j,k-1/2)}\right]}.
Corresponding to \corri\ we should have
\eqn\cotti{{\cal T}_2-<{\cal T}_2>\mapsto\int{T_{xy}+T_{yx}\over 2\pi}dxdy,}
and, assuming the correspondence extends locally
\eqn\reti{t_2(j+1/2,k)-<t_2>\mapsto{1\over 2\pi}(T_{xy}+T_{yx}).}

An formula  similar to \tliv\ can be derived for these two expressions which
from \tlii\ and \tliii\ involves commutators $[e_{2j},e_{2j-1}]$
 and $[e_{2j},e_{2j+1}]$. The result is
\eqn\resiii{2\left({\gamma\over \pi\sin\gamma}\right)^2\left([e_{2j-1},e_{2j}]+
[e_{2j},e_{2j+1}]
\right)\mapsto{1\over 2\pi}\hat{T}_{xy},}
(by reflection symmetry there is no mean value to subtract).

\subsec{Conserved quantities}

For the homogeneous model on the self dual line,
the row to row transfer matrices $\hat{\tau}(u)$ do not form a commuting family
\B\ .
 To obtain
a commuting family it is necessary, like in the Ising case \B\ ,
 to consider diagonal to diagonal transfer matrices \ref\M{
P.Martin, ``Potts model and related problems in statistical mechanics'', World
Scientific, Singapore (1989).},  (see figure 3)
\eqn\tod{\hat{\tau}_D=\prod_{j=1}^LX_{2j}X_{2j+1}.}
The matrix $\hat{\tau}_D(u)$ can then be considered as a generating function
for an
infinite family of commuting hamiltonians. The first few follow from
$$
\ln\left(Q^{-L/2}\epsilon_y^L\hat{\tau}_D(u)\right)={u\over\sin\gamma}
\sum_{j=1}^{2L}e_j+
{u^2\over2\sin^2\gamma}\sum_{j=1}^{2L}[e_j,e_{j+1}]
$$
\eqn\hi{+{u^3\over 3\sin^3\gamma}\sum_{j=1}^{2L}[e_j,[e_{j+1},e_{j+2}]]
+\cos\gamma(e_je_{j+1}+
e_{j+1}e_j)+O(u^4).}
Defining generally
\eqn\geneh{\ln\left(Q^{-L/2}\epsilon_y^L
\hat{\tau}_D(u)\right)=\sum_{n=1}^\infty\left({u\over\sin\gamma}\right)^n
{1\over n!}\hat{h}^{(n)},}
it is possible to get all the $\hat{h}^{(n)}$ by using a ladder operator.
 Introduce
\eqn\ladd{\hat{L}=\sum_{j=1}^\infty je_j,}
then  the generic term of $\hat{h}^{(n)}$ is obtained by computing the
generic term in $[\hat{h}^{(n-1)},\hat{L}]$. This follows from a simple
 generalization
of the arguments in  \ref\T{H..B.Thacker, Physica 18D (1986) 348.}
 (the statement has indeed to be given for generic
terms because there are boundary effects for a finite chain).

One the other hand we have already explained how
the spectral parameter is related to the
anisotropy angle $\theta$ \ss\ in the continuum \appi\ . As a result, consider
the transfer matrix ${\cal P}\hat{\tau}_D$ where ${\cal P}$ is the translation
operator
of half the diagonal of an elementary rectangle.
For a conformal field theory with the same geometry  (figure 4)
we would have the evolution operator (at imaginary time)
\eqn\evo{\hat{U}=\exp\left[-{2\over 2\pi}\int_{0}^{L}
\left(\sin\theta\hat{T}_{vv}+
\cos\theta\hat{T}_{uv}\right)dv\right],}
where we have rescaled by a factor 2 due to the  diagonal
geometry, $u,v$
are axis rotated by $\pi/4$ with respect to $x,y$. Therefore we expect
\eqn\cltr{{\cal P}\hat{\tau}_D(u)\mapsto \exp\left\{-{\pi\over L}\left[
\sin\theta\left(L_0+\bar{L}_0-{c\over 12}\right)+i\cos\theta
\left(L_0-\bar{L}_0\right)\right]\right\}.}
 Expanding and comparing with \geneh\
 we have
\eqn\clhi{\hat{h}^{(2n-1)}\mapsto\left({\pi\sin\gamma\over\gamma}\right)^{2n-1}
(-1)^{n}{\pi\over L}\left(
L_0+\bar{L}_0-{c\over 12}\right),}
and
\eqn\clhii{\hat{h}^{(2n)}\mapsto
\left({\pi\sin\gamma\over\gamma}\right)^{2n}(-1)^{n+1}
{\pi\over L}i\left(
L_0-\bar{L}_0\right).}
These results agree in the case $n=1$ with \resii\ and \resiii\ .

\subsec{Remarks}

Let us just notice that
\eqn\eeee{
\varepsilon(j+1/2,k)\propto\left[{e^{K_x(j+1/2,k)}-1\over e^{K_x(j+1/2,k)}}
\delta(\sigma_{jk},
\sigma_{j+1,k})+{e^{K_y(j,k+1/2)}-1\over
e^{K_y(j,k+1/2)}}\delta(\sigma_{jk},
\sigma_{j,k+1})\right],}
is invariant in $x\leftrightarrow y$ and odd in duality and is a reasonable
candidate for the energy operator in the Q state Potts model. $\varepsilon$
being odd in duality implies that all its $n$ point functions with
$n$ odd vanish, a result in agreement with the known result that
correlation functions of
$\varepsilon$ in the lattice model coincide at large distance with
those of the operator $\phi_{12}$ \DF\ in the corresponding
conformal field theory (where $\phi_{rs}$ denotes the primary field of
conformal
weight $h_{rs}$).

\subsec{Other representations of the Temperley-Lieb algebra}

Besides the Potts model, several other statistical models are obtained
by choosing different representations of the Temperley-Lieb algebra.  Such
models
include the 6 vertex model \B\ , the loop models \M\ , the ADE restricted solid
on solid models \ref\P{V.Pasquier,
Nucl. Phys. B285 (1987) 162.}.
The operator expressions like \tliv\   can still be defined in these
other models, and we believe that
they correspond to their stress energy tensors. In fact choosing
different representations of the Temperley Lieb algebra will simply  correspond
to
choosing
different representations of the Virasoro algebra.

\newsec{The scaling limit: generalities}

In this section we discuss in what sense $t_1$ can be considered a lattice
approximation
for $T_{xx}$ and define a scaling limit into which the Virasoro commutation
relations will be observed, hence giving a meaning to the notation $\mapsto$ of
the
previous section.
\subsec{The Ising model}

The  representation of the Temperley Lieb algebra corresponding to the Ising
model
 is obtained by the quotient \M\  \ref\CE{A.Connes, D.Evans, Comm.Math.Phys.
121 (1989)}\S\
\eqn\quo{e_je_{j+1}+e_{j+1}e_j-\sqrt{2}(e_j+e_{j+1})+1=0,}
which can be parametrized by
\eqn\param{e_j={1\over\sqrt{2}}+i\sqrt{2}\Gamma_j\Gamma_{j+1},}
where
\eqn\ferm{\{\Gamma_i,\Gamma_j\}=\delta_{ij}.}
The periodic RSOS model corresponds to antiperiodic fermions.
It is easy to see then that all the $\hat{h}^{(2n+1)}$ in \hi\ are linear
combinations
of terms of the form $\Gamma_j\Gamma_{j+2p+1}$ with $p\leq n$ while the
 $\hat{h}^{(2n)}$ are combinations of terms of the form
$\Gamma_j\Gamma_{j+2p}$, with
$p\leq n$. It is  therefore enough
to study the hamiltonians
\eqn\newham{\hat{\hbox{H}}^{(p)}\equiv (-1)^p{i\over 2p}\sum_{j=1}^{2L}
\Gamma_j\Gamma_{j+p}.}
One has in particular
\eqn\parti{\hat{\hbox{H}}^{(1)}={L\over 2}-{1\over 2\sqrt{2}}\hat{h}^{(1)},}
and
\eqn\partii{\hat{\hbox{H}}^{(2)}={1\over 8i}\hat{h}^{(2)}.}
We shall more generally consider the quantities
\eqn\hamf{\hat{\hbox{H}}^{(p)}(f)\equiv(-1)^p{i\over
2p}\sum_{j=1}^{2L}f\left[j+{p\over 2}\right]
\Gamma_j\Gamma_{j+p},}
where $f$ is a $C^\infty$ function of period one and we have used the short
hand notation
\eqn\shorthand{f\left[j+{p\over 2}\right]\equiv
 f\left({j+{p\over 2}\over 2L}\right).}
The function $f$ has  Fourier representation
\eqn\fourier{f(x)=\sum_lf_l e^{2i\pi lx},}
while we set
\eqn\fourieri{\Gamma_j={1\over\sqrt{2L}}\sum_{k\in {1\over 2}+Z_{2L}}
\zeta^{jk}\Psi_k,}
with
\eqn\zzi{\zeta\equiv e^{i\pi/L},}
and thus
\eqn\comii{\{\Psi_k,\Psi_{k'}\}=\delta_{k+k'}.}
By elementary computation one finds
\eqn\eqqi{\eqalign{\hat{\hbox{H}}_s^{(p)}&\equiv\hat{\hbox{H}}^{(p)}-<\hbox{u},
\hat{\hbox{H}}^{(p)}\hbox{u}>\cr
&=(-1)^{p+1}{1\over 2p}\sum_{l} f_l\left[\sum_{k,k',q}
(-1)^{pq}\sin\left({p\pi\over
2L}(k'-k)\right):\Psi_k\Psi_{k'}:\delta(k+k'+l,2qL)\right]\cr},}
with
\eqn\eqqqi{<\hbox{u},
\hat{\hbox{H}}^{(p)}\hbox{u}>=-{f_0\over p}\delta(\hbox{p odd})
{1\over 2\sin p\pi/2L},}
where the normal order is defined by
\eqn\normord{\eqalign{:\Psi_j\Psi_k:&=\Psi_j\Psi_k,\ k\geq j\cr
&=-\Psi_k\Psi_j,\ k<j\cr},}
and the ground state $|\hbox{u}>$ is annihilated by $\Psi_k$ with $k>0$.
 In this section the metric is such that $|\hbox{u}>$ has norm one and
$\Psi^{(+)}_k=
\Psi_{-k}$ (coinciding with the natural metric for Ising spins).  One has
\eqn\fermcorr{<\hbox{u},\Gamma_j\Gamma_k\hbox{u}>={1\over
4iL}{(-1)^{j-k}-1\over
 \sin{\pi\over 2L}(j-k)}.}

In this  case, what we want to call the  scaling limit  is easy to understand.
First
one restricts to   states $|\hbox{v}>$
such that
\eqn\scal{\Psi_k|\hbox{v}>=0,\ k=k_0+{1\over 2},\ldots,L-k_0-{1\over 2},}
where $k_0$ is kept fixed as $L\rightarrow\infty$. Suppose now we choose the
function $f$ to have
a single Fourier component :
 $f_l=\delta_{l,n}$. Then
 one finds,
by keeping also $n$ fixed as $L\rightarrow\infty$
\eqn\viri{\hat{\hbox{H}}_s^{(p)}(e^{2i\pi nx})
\mapsto {\pi\over 2L}\sum_k\left(k+{n\over 2}\right):a_{-k}a_{k+n}:
-(-1)^p\left(k-{n\over 2}\right):b_{-k}b_{k-n}:,}
where we have set
\eqn\defabc{\eqalign{&\Psi_k\equiv b_{k}\cr
&\Psi_{L-k}\equiv a_k\cr}.}
Due to the restriction \scal\ there is no ambiguity in this definition for $L$
large enough.
After taking the limit $L\rightarrow\infty$ one takes the limit
 $k_0\rightarrow\infty$
 to obtain indeed the well known representation of the Virasoro algebra in
 the Neveu Schwartz sector (see eg \ref\KR{V.G.Kac, A.K.Raina, "Highest weight
 representations
of infinite dimensional Lie algebras", World Scientific, Singapore (1987).})
 that is
\eqn\virrepising{\eqalign{&\hat{\hbox{H}}_s^{(2p+1)}(e^{2i\pi nx})\mapsto
{\pi\over L}
\left(L_{n}+\bar{L}_{-n}\right)\cr
&\hat{\hbox{H}}_s^{(2p)}(e^{2i\pi nx})\mapsto {\pi\over L}\left(L_{n}-
\bar{L}_{-n}\right)\cr},}
the excitations near $k=0$ and $k=L$ corresponding respectively to
 the right and left
sectors. It is this  double limit process that we  call scaling limit
in the following.

Consider now the commutator
\eqn\comisi{\eqalign{[\hat{\hbox{H}}_s^{(n_1)}(f),\hat{\hbox{H}}
_s^{(n_2)}(g)]=&(-1)^{n_1+n_2+1}{1\over 8n_1n_2}
\sum_{j=1}^{2L}
f\left[j+{s\over 2}-{n_2\over 2}\right]g\left[j+{s\over 2}+{n_1\over 2}\right
]\Gamma_j\Gamma_{j+n_1+n_2}\cr
&-f\left[j+{s\over 2}+{n_2\over 2}\right]g\left[j+{s\over 2}-{n_1\over
2}\right]
\Gamma_j\Gamma_{j+n_1+n_2}\cr
&+f\left[j+{d\over 2}+{n_2\over 2}\right]g\left[j+{d\over 2}+{n_1\over
2}\right]
\Gamma_j\Gamma_{j+n_2-n_1}\cr
&-f\left[j+{d\over 2}-{n_2\over 2}\right]g\left[j+{d\over 2}-{n_1\over
2}\right]
\Gamma_j\Gamma_{j+n_2-n_1}\cr},}
where we have set
\eqn\vardef{\eqalign{&n_2+n_1\equiv s\cr
&n_2-n_1\equiv d\cr}.}
Expanding the functions to first non trivial order we find
\eqn\comisii{\eqalign{&[\hat{\hbox{H}}_s^{(n_1)}(f),\hat{\hbox{H}}_s^{(n_2)}
(g)]=
(-1)^{n_1+n_2+1}{1\over 8n_1n_2}
{1\over 2L}\sum_{j=1}^{2L}
\left(f\left[j+{s\over 2}\right]g'\left[j+{s\over 2}\right]n_1\right.\cr
&\left.-f'
\left[j+{s\over 2}
\right]
g\left[j+{s\over 2}\right]n_2\right)
\left(\Gamma_j\Gamma_{j+n_1+n_2}-<u,\Gamma_j\Gamma_{j+n_1+n_2}u>
\right)+\cr
&\left(f\left[j+{d\over 2}\right]g'\left[j+{d\over 2}\right]n_1+f'
\left[j+{d\over 2}\right]
g\left[j+{d\over 2}\right]n_2\right)\cr
&\left(\Gamma_j\Gamma_{j+n_2-n_1}-<u,\Gamma_j\Gamma_{j+n_2-n_1}u>\right)
+\ldots+<u,[\hbox{H}_s^{(n_1)}(f),\hbox{H}_s^{(n_2)}(g)]u>\cr},}
where dots stand for higher order terms in the expansion
of the functions. For each of the two terms in the right hand side of the
above sum we can apply the previous analysis.
 Replace each sum
by its (exact) representation \eqqi\ and then consider the scaling limit for
the
right hand side  as in \viri. One
gets then
\eqn\comm{\eqalign{&\left[{L\over\pi}\hat{\hbox{H}}_s^{(n_1)}(f),{L\over\pi}
\hat{\hbox{H}}_s^{(n_2)}(g)\right]=\cr
&{1\over 2i\pi}{L\over\pi}\hbox{H}_s^{(n_1+n_2)}
(f'g-fg')+\ldots+<u,\left[{L\over\pi}\hbox{H}_s^{(n_1)}(f),
{L\over\pi}\hbox{H}_s^{(n_2)}(g)\right]u>\cr}.}
The first term on the right hand side appears only in  the scaling limit: for
finite
$L$ the algebra of lattice quantities $\hat{\hbox{H}}^{(n)}_s(f)$ does not
close.
So far we have kept $L$ finite in the  second term. Dots stand for the effect
of
higher order terms in the expansion of the functions, and corrections
to the scaling limit in \eqqi, both being negligible.
 Averages of product of gamma operators are computed
using \fermcorr. We look for all contributions that do not vanish
 as $L\rightarrow\infty$. It is easily seen that there are of two types.
Keeping only the
leading order in \fermcorr\ produces a term with third order derivatives of
functions
$f,g$, all previous order giving vanishing contributions by periodicity.
Keeping the
next order in \fermcorr\ produces a term with first order derivatives
of functions $f,g$. These two types of terms are of order $L^0$. All subsequent
contributions
vanish as $L\rightarrow\infty$. One finds therefore
\eqn\centriiiii{\eqalign{&<u,\left[{L\over\pi}\hat{\hbox{H}}_s^{(n_1)}(f),
{L\over\pi}\hat{\hbox{H}}_s^{(n_2)}(g)\right]u>=\cr
&\left({1\over (2i\pi)^3}{c\over 12}\int_0^{2L} (f'''g-fg''')dv
-{1\over 2i\pi}{c\over 12}\int_0^{2L} (fg'-f'g)dv\right)\delta(n_1+n_2\hbox{
odd})
+O(L^{-1})\cr}.}

\subsec{Scaling limit}

Let us summarize our observations  in the Ising case. A first difficulty when
one deals with a lattice model is to ``follow states'' for increasing system
sizes to be able to define a limit process. This is easily done in the
 Ising model if one
characterizes  states by the fermionic  modes which are occupied. In that
case one can also, by identifying the lattice and continuum fermions,
know  which lattice state goes to which continuum state in the  large $L$
limit.
 We then have lattice operators  like $\hat{h}^{(n)}(f)$
or equivalently the $\hat{\hbox{H}}^{(n)}(f)$ which have the following
behaviour.
If one first restricts their  action
to a finite number of low lying  excited states characterized
by some parameter $k_0$ \scal\  and   let
$L\rightarrow\infty$ the
action of these operators on the selected states coincides with the
action of Virasoro generators on the states of the continuum theory which have
been identified as the limits (in the above sense) of these lattice states.
One can then let $k_0$ go to infinity to recover the complete Virasoro
action \viri. This process is what we call scaling limit of  lattice operators.
Of course the algebra of lattice quantities does not
 close for finite $L$. However, in the Ising case,  it closes once
the scaling limit is taken for  the right hand side too.
 Moreover the
central term can be computed by taking the $L\rightarrow\infty$ limit
of the central term obtained in the lattice commutator.

The definition of the scaling
limit is easy to generalize in principle. This is done
 in practice using the Bethe-ansatz. The latter provides a way of organizing
lattice states and identifying
 them through sets of integers which bear some resemblance with the
fermionic modes of the Ising case.  We will then exhibit lattice operators
whose   scaling limit will be the Virasoro generators. However one has
 to be
careful in trying to close the algebra of lattice quantities,
  because the scaling limit of a  commutator
is  not in general the commutator of the scaling limits. To close the
 algebra, we will
need to restrict also the intermediate states
 in the computation of commutators.

 We illustrate  this point in the next subsection,
 before embarking on the study of
the scaling limit for lattice Virasoro generators in the next sections.

\subsec{Scaling limits and commutators}

Suppose that we  compute the complete
commutator of  lattice quantities using the
lattice stress energy tensor . How wrong would the
result be? we show in this section that for the measure of the central charge,
 it would in fact
be surprisingly accurate for $x$ not too large.

Consider therefore the hamiltonian $\hat{h}^{(1)}$ \clhi\  and set more
generally
\eqn\anai{\hat{\hbox{H}}(f)\equiv-{\gamma\over\pi\sin\gamma}\sum_{j=1}^{2L}
e_jf[j].}

Similarly introduce
\eqn\anaii{\hat{\hbox{P}}(f)\equiv {1\over i}\left({\gamma\over\pi\sin\gamma}
\right)^2
\sum_{j=1}^{2L}
f[j+1/2][e_j,e_{j+1}].}
Consider now
\eqn\anaiii{[\hat{\hbox{H}}(f),\hat{\hbox{H}}(g)]=
\left({\gamma\over\pi\sin\gamma}\right)^2
\sum_j(f[j]g[j+1]-f[j+1]g[j])[e_j,e_{j+1}].}
By expanding the combination of functions $f,g$ one finds
%
\eqn\anaiv{\left[{L\over\pi}\hat{\hbox{H}}(f),{L\over\pi}\hat{\hbox{H}}(g)\right]=
{1\over 2i\pi}{L\over\pi}\hat{\hbox{P}}(f'g-fg')+O\left(L^{-1}\right).}
Consider now
\eqn\anav{\eqalign{&[\hat{\hbox{H}}(f),\hat{\hbox{P}}(g)]=-{1\over
i}\left({\gamma\over\pi\sin\gamma}
\right)^3
\sum_{j=1}^{2L}f[j]\left\{g[j-3/2][e_j,[e_{j-2},e_{j-1}]]
\right.\cr
&\left.+g[j-1/2][e_j,[e_{j-1},e_j]]
+g[j+1/2][e_j,[e_j,e_{j+1}]]+g[j+3/2][e_j,[e_{j+1},e_{j+2}]]\right\}\cr}.}
Let us investigate the vacuum expectation value of this quantity, denoting by
$|\hbox{u}>$
the ground state. The metric is such that $|\hbox{u}>$ has norm one and is for
instance the
natural metric in the vertex model representation (see later).
 One finds, using translation
invariance and the Temperley Lieb defining relations
\eqn\anavi{\eqalign{&<\hbox{u},[\hat{H}(f),\hat{P}(g)]\hbox{u}>=\cr
&-{1\over
i}\left({\gamma\over\pi\sin\gamma}\right)^3\left(<\hbox{u},[e_k,[e_{k+1},
e_{k+2}]]
\hbox{u}>
\sum_{j=1}^{2L}f[j](g[j+3/2]-g[j-3/2])\right.\cr
&\left.+<\hbox{u}, \left(\sqrt{Q}(e_ke_{k+1}+e_{k+1}e_k)-2e_k\right),
\hbox{u}>\sum_{j=1}^{2L}
f[j]
(g[j+1/2]-g[j-1/2])\right)\cr}.}
Consider now the case
\eqn\partii{f(x)=e^{2i\pi px},\ g(x)=e^{2i\pi qx},}
so
\eqn\intereq{\eqalign{&\hat{\hbox{H}}(f)\mapsto {\pi\over L}
\left(L_{p}+\bar{L}_{-p}
\right)\cr
&\hat{\hbox{P}}(g)\mapsto {\pi\over L}\left(L_{q}-\bar{L}_{-q}\right)\cr}.}
If the scaling limit of the commutator was the commutator of
the  scaling limit, the right
hand side of \anavi\ would be, at dominant order
\eqn\rhsconj{\left({\pi\over L}\right)^2
\delta_{p+q}\left[2p<\hbox{u},(L_0+\bar{L}_0)\hbox{u}>+{c\over
6}(p^3-p)\right].}
For this to hold we need, from \anavi\
\eqn\conji{\eqalign{&\hbox{lim}_{L\rightarrow\infty}{4L^3\over\pi^2}
\left({\gamma\over\pi\sin\gamma}
\right)^3\left[<\hbox{u},[e_j,[e_{j+1},e_{j+2}]]u>\sin{3\pi p\over
2L}\right.\cr
&\left.+<\hbox{u},\left(\sqrt{Q}(e_je_{j+1}+e_{j+1}e_j)-2e_j\right),\hbox{u}>
\sin{\pi p\over 2L}\right]\cr
&=2p<\hbox{u},(L_0+\bar{L}_0)\hbox{u}>+{c\over 6}(p^3-p)\cr}.}
By matching the $p$ and $p^3$ coefficients on the left and right hand side
we obtain two conditions
\eqn\condi{\eqalign{&\hbox{Lim}_{L\rightarrow\infty}\left(3<\hbox{u},
[e_j,[e_{j+1},e_{j+2}]]\hbox{u}>+
\sqrt{Q}<\hbox{u},(e_je_{j+1}+e_{j+1}e_j)\hbox{u}>-2<\hbox{u},e_j\hbox{u}>\right)\cr
&=0\cr},}
and
\eqn\condii{\eqalign{&\hbox{Lim}_{L\rightarrow\infty}\left(27<\hbox{u},
[e_j,[e_{j+1},e_{j+2}]]\hbox{u}>+
\sqrt{Q}<\hbox{u},(e_je_{j+1}+e_{j+1}e_j)\hbox{u}>-2<\hbox{u},e_j\hbox{u}>\right)\cr
&=-2c\pi^2
\left({\sin\gamma\over\gamma}\right)^3\cr}.}
Other conditions would be obtained by considering the $L$ dependent
corrections to $<u,u>$, which we do not discuss for the moment. From
\condi\ and \condii\ one finds
\eqn\condiii{\hbox{Lim}_{L\rightarrow\infty}<\hbox{u},[e_j,[e_{j+1},e_{j+2}]]
\hbox{u}>\equiv
[e_j,[e_{j+1},e_{j+2}]]_\infty=
-{\pi^2c\over 12}
\left({\sin\gamma\over\gamma}\right)^3,}
a condition that is in fact  independent of $\hbox{u}$ provided $\hbox{u}$
 has eigenenergy
at distance $O(1/L)$ from the ground state.
On the other hand
 we can have access to some of these  averages by using
the known expression of the free energy of the  vertex model (or the $Q$
continuous Potts
model) in the
thermodynamic limit together with the expansion \geneh\ and
the relation between the left sides of \condii\ and \condiii\ and conserved
quantities.
Explicitely one has \B\
\eqn\condiv{\hbox{Lim}_{L\rightarrow\infty} {1\over 2L}
<\hbox{u},\ln\left(Q^{-L/2}
\epsilon_y^L\hat{\tau}_D
\right)\hbox{u}>=
\sum_{n=0}^\infty {(2u)^{2n+1}\over 2(2n+1)!}I_n,}
where
\eqn\integral{I_n=\int_{-\infty}^\infty t^{2n}{\sinh (\pi-\gamma)t\over\sinh\pi
t\cosh\gamma t}dt.}
So one has in particular
\eqn\someav{\eqalign{&e_\infty=\sin\gamma I_0\cr
&[e_j,[e_{j+1},e_{j+2}]]_\infty+\cos\gamma(e_je_{j+1}+e_{j+1}e_j)_\infty=
2\sin^3\gamma I_1\cr}.}
One finds that  \condii\ and \condiii\ imply
\eqn\test{1-{6\over x(x+1)}=-{24\pi\over (x+1)^3}
{I_0\over\sin^2[\pi/(x+1)]}+{48\pi\over (x+1)^3}
I_1.}
The integrals can easily be evaluated for $x$ integer, and are expressed as
 rational functions
of trigonometric functions.  One finds that \test\ exactly holds true
for $x=1,2,3$. For $x=3$ for instance
\eqn\exemp{I_0=1+{2\over\pi},\ I_1=1+{8\over 3\pi}.}
 For $x=4$ the righthand side is $.701184$ instead of $.70$ while for $x=6$
it is $.861348$ instead of $.85714$. Clearly for $x$ small enough, the result
 is almost true.

In the general case, we conclude that the scaling  limit of the commutator
cannot be the commutator of the scaling limits. We need to be more careful to
obtain a representation of the Virasoro
algebra from the lattice models .

\newsec{Scaling limit and conjecture for the vertex representation}

\subsec{Generalities}

The simplest case to study now is the 6 vertex model \B. In the transfer
matrix formalism  we can think of the various operators so far introduced
as acting on
the space ${\cal H}_{2L}=(C^2)^{2L}$. The elementary Temperley Lieb
matrices $e_j,\ j=1,
\ldots,2L-1$ act as
\eqn\evi{e_j=q^{-1}E_{+-,+-}+qE_{-+,-+}-E_{+-,-+}-E_{-+,+-},}
in the $j^{\hbox{th}}$ and $(j+1)^{\hbox{th}}$ copies of $C^2$  and as
identity otherwise. In
the previous equation, the basis of $C^2$ is denoted by $+,-$ and $E$ are
 unit matrices. As explained in \PS\ and
\ref\MS{P.Martin, H.Saleur, ``On an algebraic
approach to higher dimensional statistical mechanics" to appear in
 Comm. Math. Phys. and ``The blob algebra and the periodic Temperley-Lieb
 algebra" to appear in Lett. Math. Phys.}
 the last Temperley Lieb generator $e_{2L}$ reads
generally
\eqn\evi{e_{2L}=q^{-1}E_{+-,+-}+qE_{-+,-+}-e^{i\varphi}E_{+-,-+}-
e^{-i\varphi}E_{-+,+-},}
where $\varphi$ is an arbitrary complex number. Mathematically the number
 $\varphi$
selects different representations of the periodic Temperley Lieb algebra
 \tldef\ .
Physically it corresponds to twisted boundary conditions for the 6 vertex
 model
or the XXZ quantum spin chain in
the hamiltonian limit. In the following we focus
on this limit. Introduce (the normalization is such that we now consider the
chain as  being of length $2L$ while it was $L$ in the above Potts model point
 of view) the hamiltonian
\eqn\hammi{\hat{\hbox{H}}_\varphi\equiv -{\gamma\over\pi\sin\gamma}
\sum_{j=1}^{2L}\left(
e_j-e_\infty\right),}
where $e_\infty$ is $\varphi$ independent and ensures that the ground state
energy has no extensive term. Eq. \hammi\  reads as well
\eqn\hamixxz{\eqalign{&
\hat{\hbox{H}}_\varphi={2\gamma\over\pi\sin\gamma}\left(\sum_{j=1}^{2L-1}
\sigma_j^x\sigma_{j+1}^x
+\sigma_j^y\sigma_{j+1}^y+{q+q^{-1}\over 2}\sigma_j^z\sigma_{j+1}^z\right.\cr
&\left.+{e^{i\varphi}
\over 2}
\sigma_1^+\sigma_{2L}^-+{e^{-i\varphi}
\over 2}
\sigma_1^-\sigma_{2L}^++{q+q^{-1}\over 2}\sigma_1^z\sigma_{2L}^z-L{q+q^{-1}
\over 4}-2Le_\infty\right)\cr}.}
Notice that $[\hat{H}_\varphi,S^z]=0$ where $S^z=\sum\sigma_i^z$.
 We can therefore decompose
\eqn\dechilb{{\cal H}_{2L}=\bigoplus_{S^z=-L}^L{\cal H}_{2L}^{S^z},}
where the sum runs over integers.

Introduce also $\hat{\hbox{P}}_\varphi$ by
\eqn\pdef{\hat{\hbox{P}}_\varphi={1\over i}
\left({\gamma\over\pi\sin\gamma}\right)^2\sum_{j=1}^{2L}
[e_j,e_{j+1}].}
The following result has been established by several non rigorous
 approaches, and
carefully checked numerically \ref\ABB{F.C.Alcaraz, M.N.Barber,
 M.T.Batchelor, Phys. Rev. Lett. 58 (1987) 771.}\ref\DFSZ{P.di Francesco,
 H.Saleur,
J.B.Zuber, J.Stat. Phys. 49 (1987) 57.}. Consider the limit
 $L\rightarrow\infty$. Introduce
two other positive numbers $T_R,T_I$ such that
\eqn\limdef{ L,T_R,T_I\rightarrow\infty,\  T_R/L\rightarrow t_R, \
T_I/L\rightarrow
t_I \hbox{ with $t_R,t_I$ finite}.}
Then
\eqn\limitone{\lim\  tr_{{\cal H}_{2L}^{S^z}}\exp (-T_I\hat{\hbox{H}}_{\varphi}
-
iT_R\hat{\hbox{P}}_\varphi)=
{1\over\eta(p)\eta(\bar{p})}\sum_{e\in Z} p^{{1\over 4}\left[(e-e_\varphi)
\alpha_++S^z\alpha_-\right]^2}\bar{p}^{{1\over 4}\left[(e-e_\varphi)
\alpha_+-S^z\alpha_-\right]^2},}
where the limit is defined in \limdef\ , $e_\varphi={\varphi\over2\pi}$,
 $p=\exp(-\pi t_I-i\pi t_R)$ and $\eta(p)=p^{1/24}\prod_{n>0}(1-p^n)$.
 This coincides with the  trace of
$$
p^{L_0-c/24}\bar{p}^{\bar{L}_0-c/24},
$$
over the space
\eqn\virrep{\bigoplus_{\alpha}F_{\alpha,\alpha_0}\otimes\bar{F}_{\bar{\alpha},
\alpha_0},}
where $F_{\alpha,\alpha_0}$ is the Gaussian Fock space built of fields of
 the form
\eqn\fock{P(\partial\phi,\partial^2\phi,\ldots)\exp i\alpha\phi,}
where $\phi$ is a chiral boson \ref\DF{V.Dotsenko, V.A.Fateev, Nucl.
 Phys. B240 (1984) 312.}\ref\F{G.Felder, Nucl. Phys. B317 (1989) 215.}.
 The sum is taken over $e\in Z$ with
\eqn\alp{\alpha(\bar{\alpha})={e-e_\varphi\over 2}\alpha_++\alpha_0+(-)
{S^z\over 2}\alpha_-.}
The  twisted stress-energy tensor reads
\eqn\stcft{T(z)=-(\partial\phi)^2+i\alpha_0\partial^2\phi,}
with
\eqn\alphs{\alpha_0={1\over 2\sqrt{x(x+1)}},\alpha_+=\sqrt{{x+1\over x}},
\alpha_-=-
\sqrt{{x\over x+1}},}
and we have parametrized $q=\exp(i\pi/(x+1))$. This system has central charge
\eqn\ccen{c=1-24\alpha_0^2=1-{6\over x(x+1)},}
and an "effective central charge" equal to one due to the negative
dimension operator $e=S^z=0,e_\varphi=0,\alpha=\alpha_0$.

The same result would hold if instead of \hammi\ and \pdef\ we considered
any of the other (properly scaled) hamiltonians obtained from \clhi\
and \clhii, call them $\hat{\hbox{H}}^{(2n+1)}_\varphi$ and
$\hat{\hbox{H}}^{(2n)}_\varphi$ respectively.

In the following it will be convenient to use  the Heisenberg algebra (see eg
\KR\ )
that follows from
a mode expansion of the chiral boson
$$
[a_n,a_m]=n\delta_{n+m},\ n,\ m\in Z.
$$
Corresponding to the vertex operators $\exp i\phi$  we have then vectors
$v_{\alpha,\alpha_0}$ with
$$
a_nv_{\alpha,\alpha_0}=0\hbox{ for }n>0,\
a_0v_{\alpha,\alpha_0}=\sqrt{2}(\alpha-\alpha_0)
 v_{\alpha,
\alpha_0},
$$
and the Fock space  $F_{\alpha,\alpha_0}$ is built as
\eqn\fockdef{F_{\alpha,\alpha_0}=\bigoplus_{k=0}^\infty\bigoplus_{1\leq
n_1\leq\ldots\leq
 n_k}Ca_{-n_1}\ldots a_{-n_k}v_{\alpha,\alpha_0}.}
Then  $F_{\alpha,\alpha_0}$ is a Virasoro module with generators
\eqn\viri{\eqalign{&L_n={1\over 2}\sum_{k=-\infty}^\infty :a_{n-k}a_k:-
\sqrt{2}\alpha_0na_n,\ n\neq 0\cr
&L_0=\sum_{k=1}^\infty :a_{-k}a_k:+{1\over 2} a_0^2-\alpha_0^2\cr},}
where normal order is defined by putting the largest index on the right.
The space  $F_{\alpha,\alpha_0}$ is graded by eigenspaces of $L_0$
%
\eqn\graded{F_{\alpha,\alpha_0}=\bigoplus_{n=0}^\infty\left(F_{\alpha,\alpha_0}\right)_n.}
The dimension of the eigenspace $\left(F_{\alpha,\alpha_0}\right)_n$ with $L_0$
eigenvalue
$\alpha^2-2\alpha\alpha_0+n$ is $p(n)$ the number of partitions of $n$.

We define the positive definite  hermitean form  $<,>$
on the Fock space $F_{\alpha,\alpha_0}$ contravariant with respect to
\eqn\herm{a_n^+\equiv a_{-n},}
where by convention the vector $v_{\alpha,\alpha_0}$ has norm one and
 $v_{\alpha,\alpha_0},v_{\alpha',\alpha_0}$
are orthogonal for $\alpha\neq\alpha'$.
Notice that with this definition the  representation
of the Virasoro algebra in $F_{\alpha,\alpha_0}$ is not unitary since
\eqn\virdag{L_{n}^+={1\over 2}\sum_{k=-\infty}^\infty
:a_{-n-k}a_k:-\sqrt{2}\alpha_0
 na_{-n}\neq L_{-n}.}
Observe that with the natural scalar product in ${\cal H}_{2L}^{S^z}$, the
Temperley-Lieb matrices \evi\ obey also $e_j^+\neq e_j$.

\subsec{Scaling limit and Bethe ansatz}

As stressed in section 3 one must be careful with
the analysis of
 ${\cal H}_{2L}^{S^z}$. First there does not seem to be a simple
way to imbed
 ${\cal H}_{2L}^{S^z}$ in \virrep\ . Hence it is difficult to
 follow states when
$L$ increases and define a suitable limiting process. Also
 ${\cal H}_{2L}^{S^z}$ is
obviously "too big" \ref\TF{L.A.Takhtajan, L.D.Fadeev,
 Phys. Lett. A85 (1981) 375.}. To proceed further we
 use the fact that the
hamiltonian $\hat{\hbox{H}}_\varphi$ is diagonalizable by Bethe ansatz
\ref\ABB{F.C.Alcaraz, M.N.
Barber, M.T.Batchelor, Phys. Rev. Lett. 58 (1987) 771.}
\ref\ABBI{F.C.Alcaraz, M.N.
Barber, M.T.Batchelor, J.Stat. Phys.  182 (1988) 280.}. The eigenstates
(we restrict to $S^z>0$ due
to the symmetry $(S^z,\varphi)\leftrightarrow (-S^z,-\varphi)$)
 are sums of plane waves
with momenta $\{k_j\}$ $(j=1,\ldots,L-S^z)$  solutions of the
set of coupled Bethe equations
\eqn\bethe{2Lk_j=2\pi I_j+\varphi-\sum_{l\neq j}\Theta(k_j,k_l),}
where $\Theta$ is the usual kernel of the XXZ chain \ref\G{M.Gaudin, ``La
fonction d'onde de Bethe'', Masson.}
and $\{I_j\}\in Z(Z+{1\over 2})$ for $L-S^z$ odd (even) and are all
 distinct. The corresponding
 eigenenergy reads
\eqn\eigen{E=-{\gamma\over\pi\sin\gamma}\left[2\sum_{j=1}^{L-S^z}
\left({q+q^{-1}
\over 2}+\cos k_j\right)-2Le_\infty\right],}
and the momentum obtained by considering properties of the wave function
under translations reads
\eqn\realmom{-{\cal P}={\pi\over L}\sum_{j=1}^{L-S^z} I_j+{\varphi\over 2L}
(L-S^z).}
We shall characterize the eigenstates and their
corresponding  eigenenergies  by the associated set of numbers $\{I_j\}$.
The ground state in
every ${\cal H}_{2L}^{S^z}$ is obtained by choosing the maximally packed set
 of $I_j$,
symmetrically distributed around the origin : the associated
 eigenvalues of
 $\hat{\hbox{H}}_\varphi$ and $\hat{\hbox{P}}_\varphi$
 reproduce in the limit \limdef\ the conformal weights of
$v_{\alpha,\alpha_0}$ with $\alpha=-{e_\varphi\over 2}\alpha_++\alpha_0+
{S^z\over 2}
\alpha_-$. More precisely one has
$$
\hbox{Lim}_{L\rightarrow\infty} {L\over\pi}E(\{I_j\})=
(\alpha^2-2\alpha
\alpha_0)+(\bar{\alpha}^2-2\bar{\alpha}
\alpha_0)-
{c\over 12}.
$$

 By shifting as a  whole
the set $\{I_j\}$ by $e$ units (to the left or
to the
right depending on the sign of $e$)  one obtains an excited state whose
associated
eigenvalues
reproduce in the limit \limdef\ the conformal weights of
$v_{\alpha,\alpha_0}\otimes\bar{v}_{\bar{\alpha},\alpha_0}$ with
 $\alpha(\bar{\alpha})={e-e_\varphi\over 2}\alpha_++\alpha_0+(-){S^z\over 2}
\alpha_-$. Creating "holes"  in the (shifted) set of integers
can lead to
further excited states whose eigenvalues reproduce in the limit \limdef\
the conformal
weights of descendents in $F_{\alpha,\alpha_0}\otimes\bar{F}_{\bar{\alpha},
\alpha_0}$.
 Call
$\{I_j^0\}$ the particular distribution that leads to $v_{\alpha,\alpha_0}
\otimes\bar{v}_{\bar{\alpha},\alpha_0}$.
 Call $\{I_j\}$
the distribution obtained by making some holes; then the eigenvalues of the
associated state approaches the ground state as
 $1/L$ provided the set $\{I_j-I_j^0\}$ has a finite, $L$
independent,  subset of non vanishing integers .
In the simple case
when $\varphi=0$ and  all the momenta
$k_j$ in
\bethe\ are real, the corresponding conformal weights read
\eqn\hshifI{\eqalign{&h=h_{\alpha,\alpha_0}+{1\over 2}
\sum\left[ (I_j-I_j^0)+|I_j-I_j^0|\right]\cr
&\bar{h}=\bar{h}_{\alpha,\alpha_0}-{1\over 2}\sum \left[(I_j-I_j^0)-
|I_j-I_j^0|\right]\cr},}
so the left and right excitations occur for jumps of positive and negative
$I_j$.

If we characterize the states by the integers $\{I_j\}$ we can put
eigenstates for various $L$ in correspondence and define in that fashion
 a limit process, which we also denote by the symbol $\mapsto$. We can also
simply isolate the eigenenergies that will contribute to \limitone\ .
The
associated set $\{I_j\}$ has to differ from a symmetric, maximally
 packed distribution
by finite global shifts and finite numbers of holes, these
deviations being moreover $L$ independent
 \ABBI\ \ref\K{M.Karowski, Nucl. Phys. B300 (1988) 473.}.
 This picture becomes transparent
in the case $q=i$ where the $\Theta$ term in \bethe\ vanishes and the spectrum
 is the one of a
free theory. See also the Ising model of the previous section.
We sometimes call  states characterized by such distributions of
integers: scaling states. We shall talk also about the limit of these states,
meaning that we consider the family of states  characterized by similar
patterns
of $\{I_j\}$ as $L$ increases. Usually the $L$ dependence will not be written
explicitely, in particular in the numerical results of section 5.

Note that we implicitely assumed that all eigenenergies could be found
from the Bethe
ansatz. This is generally believed to be true. For special values
 of $q$ or $\varphi$
this may be true up to degeneracies. But in such cases, using the
 known additional symmetries
of the hamiltonian \PS\ , the correct multiplicities can be recovered.

Note that in practice, when the hamiltonian is diagonalized numerically,
 it is easy to ``follow''  the eigenenergies and their
associated eigenstates by simple order. Comparing the $k^{\hbox{th}}$
 ($k$ fixed)
eigenenergies for increasing  values of  $L$, is a reliable limit process.
Notice also that  the momentum can be measured without finite size
corrections from ${\cal P}$, which makes the identification of terms in
\limitone\
 easier.

\subsec{Lattice Virasoro generators}

We now state the main conjecture of this section. Introduce the
 lattice quantities
\eqn\lattvir{l_n={L\over 2\pi}\left\{-{\gamma\over\pi\sin\gamma}\sum_{j=1}^{2L}
e^{inj\pi/L}\left(e_j-e_\infty
+{i\gamma\over\pi\sin\gamma}[e_j,e_{j+1}]\right)\right\}+{c\over
24}\delta_{n,0},}
and
\eqn\lattvirbar{\bar{l}_{-n}={L\over 2\pi}\left\{-{\gamma\over\pi\sin\gamma}
\sum_{j=1}^{2L}
e^{inj\pi/L}\left(e_j-e_\infty-{i\gamma\over\pi\sin\gamma}
[e_j,e_{j+1}]\right)
\right\}+{c\over 24}\delta_{n,0}.}
Consider the following double limit process. Choose a value of $\alpha$ and
$\bar{\alpha}$ and a pair of
integers $N,\bar{N}$.
Using the characterization of Bethe eigenstates by integers, or
simply by ordering the
eigenenergies, select for every $L$ the set of
$$
\sum_{n=0}^Np(n)\sum_{\bar{n}=0}^{\bar{N}}p(\bar{n}),
$$
eigenstates associated with gaps gaps which, after multiplication by $T_R$
reproduce in the limit $L\rightarrow\infty$ the conformal weights
 $h_{\alpha,\alpha_0}+n$
and $\bar{h}_{\alpha,\alpha_0}+\bar{n}$ with $n\leq N,\bar{n}\leq\bar{N}$
 (for $L$ large enough
but finite all these states will be present: we implicitely suppose we
 always are in such
 a situation). Consider the action of the generators $l_n$ (resp.
$\bar{l}_n$) restricted to these
states. We conjecture that it furnishes in the limit $L\rightarrow\infty$ a
representation
 of $\hbox{Vir}$
 (resp. $\bar{\hbox{Vir}}$) restricted to
$$
\bigoplus_{n=0}^N\bigoplus_{\bar{n}=0}^{\bar{N}}\left(F_{\alpha,\alpha_0}
\right)_n\otimes\left(\bar{F}_{\bar{\alpha},\alpha_0}\right)_{\bar{n}},
$$
that is  more precisely a representation of $P\hbox{Vir}P$ (resp.
$P\bar{\hbox{Vir}}P$ where $P$ is the projector on the
above subspace).  Moreover we conjecture that the  natural scalar product in
${\cal H}_{2L}$
(with $<+,+>=<-,->=1$ and $<+,->=<-,+>=0$) coincides in
the $L\rightarrow\infty$ limit
with the hermitean form defined previously  in terms of Heisenberg algebra.

\newsec{Numerical and analytic checks}

A first remark is that
\eqn\momcom{[l_n,{\cal P}]=nl_n,\ [\bar{l}_n,{\cal P}]=-n\bar{l}_n,}
so the lattice $l_n,\bar{l}_n$ have non vanishing matrix elements only between
eigenspaces of the lattice momentum ${\cal P}$ whose eigenvalues differ
 by $\pm n$. This
is of course the same result as what is expected in the $L\rightarrow\infty$
limit.

\subsec{Level one}

To start consider the ground state in each sector ${\cal H}_{2L}^{S^z}$; call
 it u.
 It should
correspond to $v_{\alpha,\alpha_0}$ with $\alpha=-{e_\varphi\over 2}\alpha_++
\alpha_0+
{S^z\over 2}\alpha_-$.
In each sector  ${\cal H}_{2L}^{S^z}$ consider also the first (normalized)
excited state with momentum $+1$, which should correspond, up to a
proportionnality
 factor, to $a_{-1}v_{\alpha,\alpha_0}$; call it v. In the
continuum theory one has
\eqn\numresi{\left<L_{-1}v_{\alpha,\alpha_0},L_{-1}v_{\alpha,\alpha_0}\right>=
\left<v_{\alpha,\alpha_0},L_{-1}^+ L_{-1}v_{\alpha,\alpha_0}\right>=
2\alpha^2=2\left(\alpha_0-{e_\varphi\over 2}\alpha_++{S^z\over 2}\alpha_-
\right)^2,}
and thus
\eqn\numresii{
\left<L_{-1}v_{\alpha,\alpha_0},L_{-1}v_{\alpha,\alpha_0}\right>^{1/2}=
\sqrt{2}\left|\alpha_0-{e_\varphi\over 2}\alpha_++{S^z\over 2}\alpha_-\right|.}
We thus checked numerically in a variety of cases that
\eqn\nri{\left<\hbox{v},l_{-1}\hbox{u}\right>\rightarrow \sqrt{2}\left
|\alpha_0-{e_\varphi\over 2}\alpha_++
{S^z\over 2}\alpha_-\right|,}
and similarly
\eqn\nrii{\left<\hbox{u},l_1\hbox{v}\right>\rightarrow -\sqrt{2}\left|
\alpha_0+{e_\varphi\over 2}\alpha_+
-{S^z\over 2}\alpha_-\right|.}
We also checked that
\eqn\nriii{\left<\hbox{v},\bar{l}_{1}\hbox{u}\right>\rightarrow 0,\
\left<\hbox{u},\bar{l}_{-1}\hbox{v}\right>\rightarrow 0.}
A special value is $e_\varphi=-2\alpha_0\alpha_-$ and $S^z=0$
 for which \nri\ becomes
\eqn\supeqi{
\left<\hbox{v},l_{-1}\hbox{u}\right>\rightarrow 0,}
and $e_\varphi=2\alpha_0\alpha_-$ and $S^z=0$ for which \nrii\ becomes
\eqn\supeqii{
\left<\hbox{u},l_1\hbox{v}\right>\rightarrow 0.}
In fact the first result holds even for finite $L$. It is a consequence of the
representation theory of the periodic Temperley Lieb algebra \MS; this is
explained in subsection 5.7. Some numerical results for the foregoing
quantities are given in tables 1,2,3,4.

Recall finally that  from \momcom\  we have exactly
\eqn\supeqiii{
\left<\hbox{v},\bar{l}_{\pm n}\hbox{u}\right>=0,\ \left<\hbox{u},l_{\pm n}
\hbox{v}
\right>=0,}
for  $n>1$.

\subsec{Level two}

Consider for simplicity the case  ${\cal H}_{2L}^{S^z=0}$ only,
and identify the two orthonormal
excited states above the ground state u, with momentum two. For any $x$
it is easy to
find one of them by numerical solution of the Bethe ansatz equations.
 It corresponds to
integers $\{I_j\}$ obtained from $\{I_j^0\}$ by shifting the two rightmost
integers by one unit. We call it $\hbox{w}_1$. We had difficulties
finding the other one
by numerical solution of \bethe\  (which should be obtained by
shifting the the rightmost integer by two units to the right)
so we used direct diagonalization (Lanczos algorithm
\ref\Henkel{P.Christe, M.Henkel,
``Introduction to conformal invariance and its
 applications to critical phenomena'', preprint UGVA/DPT1992/11-794.}).
Call this
other state  $\hbox{w}_2$.
In the large $L$ limit
we expect
\eqn\tatai{|\hbox{w}_1>\mapsto {1\over \sqrt{2x^2+2y^2}}
\left(xa_{-1}^2+ya_{-2}\right) |v_{\alpha,\alpha_0}>,}
where $x,y$ are coefficients to be determined. To help us determine
the coefficients $x,y$ we made the following observations.
Consider first the  the case $e_\varphi=0$. We found then
$$
<\hbox{u},l_2\hbox{w}_1>=-<\hbox{w}_1,l_{-2}\hbox{u}>,
$$
this result being true for any finite $L$, and a fortiori
 for $L\rightarrow\infty$. From this we deduce
\eqn\tataii{x=0 \hbox{ if }e_\varphi=0.}
Hence
\eqn\tataiii{|\hbox{w}_1>\mapsto
 {1\over\sqrt{2}}
a_{-2}|v_{\alpha,\alpha_0}>,
 |\hbox{w}_2>\mapsto {1\over\sqrt{2}}a_{-1}^2|v_{\alpha,\alpha_0}>,}
and
\eqn\tataiv{\eqalign{&<\hbox{w}_1,l_{-2}\hbox{u}>\rightarrow 4\alpha_0\cr
&<\hbox{w}_2,l_{-2}\hbox{u}>\rightarrow {1\over\sqrt{2}}\cr}.}
We also observed that
and
$$
<\hbox{u},l_2\hbox{w}_2>=<\hbox{w}_2,l_{-2}u>,
$$
holds exactly for finite $L$ too.
Corresponding numerical results are shown in table 5.

Similarly  in the case $e_\varphi=-2\alpha_0\alpha_-$ we found
$$
<\hbox{w}_1,l_{-2}\hbox{u}>=0,
$$
this result being true for finite $L$. From this we deduce
\eqn\tatav{x=-2\sqrt{2}\alpha_0 y\hbox{ if }
e_\varphi=-2\alpha_0\alpha_-,}
and then
\eqn\tatavi{|\hbox{w}_1>\mapsto {1\over\sqrt{2+16\alpha_0^2}}\left(-
2\sqrt{2}\alpha_0a_{-1}^2+a_{-2}\right) |v_{\alpha,\alpha_0}>,}
and
\eqn\tatavii{|\hbox{w}_2>\mapsto {1\over\sqrt{2+16\alpha_0^2}}
\left(a_{-1}^2+2\sqrt{2}
\alpha_0
a_{-2}\right)|v_{\alpha,\alpha_0}>={2\over\sqrt{2+16\alpha_0^2}}L_{-2}|
v_{\alpha,\alpha_0}>.}
So we get the results
\eqn\tataviii{\eqalign{&<\hbox{u},l_{2}\hbox{w}_1>\rightarrow -{8\alpha_0\over
\sqrt{1+8\alpha_0^2}}\cr
&<\hbox{u},l_{2}\hbox{w}_2>\rightarrow{1-24\alpha_0^2\over
\sqrt{2+16\alpha_0^2}}\cr
&<\hbox{w}_2,l_{-2}\hbox{u}>\rightarrow\sqrt{{1+8\alpha_0^2\over 2}}\cr}.}
Corresponding numerical results are shown in table 6.

We can also consider action of $l_{2}$. An especially interesting quantity
is then (still for $e_\varphi=-2\alpha_0\alpha_-$)
\eqn\tataix{<\hbox{u},l_{2}\hbox{w}_1><\hbox{w}_1,l_{-2}\hbox{u}>+
<\hbox{u},l_{2}\hbox{w}_2><\hbox{w}_2,l_{-2}\hbox{u}>\rightarrow {c\over 2}.}
Numerical results are given in table 7,8. They actually provided very
good estimates of the central charge.

We can also consider action of $l_{-1}^2$. For instance,
 for $e_\varphi=0$,
\eqn\tatax{\eqalign{&<\hbox{w}_1,l_{-1}\hbox{v}><\hbox{v},l_{-1}\hbox{u}>
\rightarrow 2\alpha_0\cr
&<\hbox{w}_2,l_{-1}\hbox{v}><\hbox{v},l_{-1}\hbox{u}>
\rightarrow 2\sqrt{2}\alpha_0^2\cr},}
while for $e_\varphi=-2\alpha_0\alpha_-$,
\eqn\supptata{<\hbox{u},l_{1}\hbox{v}><\hbox{v},l_{1}\hbox{w}>
\rightarrow 4\alpha_0\sqrt{1+8\alpha_0^2}.}
Corresponding numerical results are in table 9.

Also, at level two we can also consider action of $l_{-1}$ and
 $\bar{l}_{-1}$. For instance consider the eigenstate $|\hbox{w}>$ with
 vanishing momentum, associated with a gap that reproduces the
conformal weights $h=\bar{h}=1$ in the limit \limitone. One
expects then
\eqn\supptatai{<\hbox{w},\bar{l}_{-1}\hbox{v}><\hbox{v},l_{-1}\hbox{u}>
\rightarrow 2\alpha_0^2.}
Numerical results for this quantity are given in table 10.

At level two we can finally test numerically a non-trivial case of
degenerescence. Consider, with $e_\varphi=-2\alpha_0\alpha_-$ and
 $S^z=0$ the state $|\hbox{u}'>$ of conformal
weight $h_{21}$.
This state is degenerate
 at level two. From the Coulomb gas mapping we expect that

\eqn\supptataii{<\hbox{w}',l_{-2}\hbox{u}'>-{3\over 2(2h_{21}+1)}
<\hbox{w}',l_{-1}
\hbox{v}'><\hbox{v}',l_{-1}\hbox{u}'>,}
will vanish for large systems, where $|\hbox{v}'>$ and $|\hbox{w}'>$
are the appropriate excited states.
 Numerical results are shown in table 11 and converge very well to zero.

\subsec{Level three}

Consider now  ${\cal H}_{2L}^{S^z=0}$ and identify the three orthonormal
excited
eigenstates states above the ground state, with momentum three. Call them
$\hbox{y}_{1},\hbox{y}_{2},\hbox{y}_{3}$.
In the large $L$ limit we expect
\eqn\wmi{\eqalign{|\hbox{y}_1>&\mapsto
{1\over \sqrt{6c_1^2+2c^{'2}_{1}+3}}(c_1 a_{-1}^{3}+c^{'}_1
a_{-1}a_{-2}+a_{-3})
|v_{\alpha,\alpha_0}>\cr
|\hbox{y}_2>&\mapsto{1\over\sqrt{6c_{2}^{2}+2+3c_{2}^{'2}}}(c_2 a_{-1}^{3}+
a_{-1}a_{-2}+c^{'}_2 a_{-3})|v_{\alpha,\alpha_0}>\cr
|\hbox{y}_3>&\mapsto{1\over\sqrt{6c_3^2+2c^{'2}_{3}+3}}(c_3 a_{-1}^{3}+
c^{'}_3 a_{-1}a_{-2}+a_{-3})|v_{\alpha,\alpha_0}>.\cr}}
To determine  the  unknown coefficients, we use orthogonality of the
 states, which implies
\eqn\wmii{\eqalign{6c_1 c_2+2c_{1}^{'}+3c_{2}^{'}&=0\cr
           6c_3c_2+2c_3^{'}+3c_2^{'}&=0\cr
           6c_1c_3+2c_1^{'}c_3^{'}+3&=0,\cr}}
and the numerical obervation that for $e_{\varphi}=0$,
\eqn\wmiii{\eqalign{<\hbox{y}_1,l_{-3}u>&=-<\hbox{u},l_{3}\hbox{y}_1>\cr
           <\hbox{y}_2,l_{-3}\hbox{u}>&=<\hbox{u},l_3\hbox{y}_2>\cr
           <\hbox{y}_3,l_{-3}\hbox{u}>&=-<\hbox{u},l_3\hbox{y}_3>.\cr}}
{}From the above, we deduce that all the coefficients
 vanish except for $c_1,c_3$ which are related by
\eqn\wmiv{c_1c_3=-1/2.}
In addition, the above matrix elements of $l_{-3}$ are now given by
\eqn\wmv{\eqalign{<\hbox{y}_1,l_{-3}\hbox{u}>&=
{3\sqrt{6}\alpha_{0}\over\sqrt{1+2c_1^2}}\cr
           <\hbox{y}_2,l_{-3}\hbox{u}>&=\sqrt{2}\cr
           <\hbox{y}_3,l_{-3}\hbox{u}>&=
{6\sqrt{3}c_1\alpha_0\over\sqrt{1+2c_1^2}}.\cr}}
There does not seem to be any way to determine the value of the unknown $c_1$.
 That
$c_1c_3\neq 0$ shows that the eigenstates of finite chains do not  converge in
general
to pure monomials in the Heisenberg algebra  as was the case at level two.
Even if $c_1$ is not known observe from \wmv\ that
\eqn\wmvi{\left(<\hbox{y}_1,l_{-3}\hbox{u}>^2+
<\hbox{y}_3,l_{-3}\hbox{u}>^2\right)^{1/2}=3\sqrt{6}\alpha_0,}
a result in dependent of $c_1$ that can therefore lead to numerical study.
Numerical
results are shown in table 12.

Similarly for $e_{\varphi}=-2\alpha_0\alpha_-$, one of the
 three excited states $y_2$ corresponds,
 in the continuum limit, to the level 3 descendant of the vacuum
 $|v_{\alpha,\alpha_0}>$ and is given by
\eqn\wmvii{\eqalign{|\hbox{y}_2> &\mapsto {1\over\sqrt{2+24\alpha_{0}^{2}}}
l_{-3}|v_{\alpha,\alpha_0}>\cr
       &={1\over\sqrt{2+24\alpha_{0}^{2}}}(a_{-1}a_{-2}
+2\sqrt{2}\alpha_{0}a_{-3})|v_{\alpha,\alpha_0}>.\cr}}
While the other two excited states $y_1,y_3$ correspond,
 in the continuum limit, to  two level two descendants
 of the null states and are therefore  given by
\eqn\wmix{\eqalign{|\hbox{y}_1> \mapsto &{1\over
\sqrt{(3+54\alpha_0^{2})(1+2\lambda_{1})^2
+6({1\over 2}+8\alpha_0^{2})^2}}(l_2^{\dagger}+\lambda_{1}
l_1^{2\dagger})a_{-1}
|v_{\alpha,\alpha_0}>\cr
&={1\over \sqrt{3+36\alpha_{0}^{2}+6c_{1}^{2}}}(c_1 a_{-1}^{3}-3\sqrt{2}
\alpha_{0}a_{-1}a_{-2}+a_{-3})|v_{\alpha,\alpha_0}>\cr
|\hbox{y}_3>\mapsto &{1\over\sqrt{(3+54\alpha_0^2)(1+2\lambda_{3})^2
+6({1\over 2}+8\alpha_0^2)^2}}(l_2^{\dagger}+\lambda_{3}l_1^{2\dagger})a_{-1}
|v_{\alpha,\alpha_0}>\cr
&={1\over \sqrt{3+36\alpha_{0}^{2}+6c_{3}^{2}}}
(c_3 a_{-1}^{3}-3\sqrt{2}\alpha_{0}a_{-1}a_{-2}+a_{-3})
|v_{\alpha,\alpha_0}>\cr},}
where we have
traded the unknown $\lambda_{1(3)}$ by another  $c_{1(3)}$ for convenience.
 Orthogonality then implies that
\eqn\wmx{c_1c_3=-(1/2+6\alpha_0^{2}).}
We therefore expect
\eqn\wmxi{\eqalign{<y_1,l_{-3}u>&\longrightarrow 0\cr
           <u,l_{3}y_1>&\longrightarrow -{18\sqrt{2}\alpha_{0}\over
\sqrt{3+36\alpha_{0}^{2}+6c_{1}^{2}}}\cr
<y_2,l_{-3}u>&\longrightarrow \sqrt{2+24\alpha_{0}^{2}}\cr
           <u,l_{3}y_2>&\longrightarrow {2-48\alpha_{0}^{2}\over
\sqrt{2+24\alpha_{0}^{2}}}\cr
<y_3,l_{-3}u>&\longrightarrow 0\cr
           <u,l_{3}y_3>&\longrightarrow -{18\sqrt{2}\alpha_{0}
\over\sqrt{3+36\alpha_{0}^{2}+6c_{3}^{2}}}\cr},}
Numerical calculation shows that $<y_{1(3)},l_{-3}u>$ vanish exactly
 for finite $L$ and as in the case of $e_{\varphi}=0$,
the coefficient $c_{1}$ remains unknown. However
\eqn\wmvibis{\left(<\hbox{y}_1,l_{-3}\hbox{u}>^2+
<\hbox{y}_3,l_{-3}\hbox{u}>^2\right)^{1/2}={6\sqrt{6}\alpha_0
\over \sqrt{1+12\alpha_0^2}},}
a combination that is independent of $c_1$.  Some numerical values  are given
in table 12 and 13.

\subsec{Higher order approximations}

According to eqs. \clhi\ and \clhii\ we can form lattice approximations to the
Virasoro generators by choosing any generic term of higher order conserved
quantities. As an example we give in tables 14 and 15 measures of the central
charge from \tataix. The results are comparable to the ones obtained with the
lowest hamiltonians.

\subsec{The double limit process}

The  double limit process is very important in the conjecture.
However it plays only a small role numerically, as was already observed in the
discussion after eq. \test.  Consider for instance the lattice quantity
\eqn\exi{<l_{-1}\hbox{u},l_{-1}\hbox{u}>.}
Compare this expression to \numresi\ .
 For instance for $e_\varphi=0$ and $S^z=0$
eq. \numresi\ gives $2\alpha_0^2={1\over 2x(x+1)}$.
For $e_\varphi=-2\alpha_o\alpha_-$
eq. \numresi\ gives zero. Measures of \exi\ for
 $x=3,7$ are given in tables 16 and 17. The data
do not behave as nicely as the ones  for
\eqn\exii{<l_{-1}\hbox{u},\hbox{v}><\hbox{v},l_{-1}\hbox{u}>,}
which can even give exact results (as a consequence of Temperley-Lieb
representation theory) for finite $L$. It is very likely  that they
are not converging to  \numresi\ since they are monotonous and above \numresi\
for $L>7$. This is expected for  the reason already discussed in section 3.2:
in \exi\ all
 intermediate states contribute, including
those at large distance from the ground state. Although matrix elements
 of $l_{-1}$ between u and states different from $\hbox{v}$
are expected to vanish in the limit $L\rightarrow\infty$, the sum of all
the corresponding
contributions may well remain finite.

\subsec{Numerical Comments}

In general, we think that the numerical agreement with the conjecture is very
 good.
The measure of $c$ from \tataix\ in particular provides results of the same
precision as direct study of finite size effects for the ground state. It so
happens
that for a particular quantity and a particular value of $x$ the agreement is
not perfect (see eg table 1).
This however can usually be attributed to data being non monotonic, hence
difficult to extrapolate. But for a given value of $x$, most tests give good
results.

The difference between limit of commutator and commutator of limits seems
obeservable  numerically, but is rather weak.

\subsec{Some analytic checks from Temperley-Lieb representation theory}

Some of the conjectured results actually hold exaclty for finite $L$ and
can be established by using the representation theory of the periodic
 Temperley-Lieb algebra, in particular the analysis of the vertex
 model representation in \MS. Consider first
the case $x$ irrational. Then
for generic $\varphi$, the space ${\cal H}_{2L}^{S^z}$ provides an
irreducible
representation of the periodic Temperley-Lieb algebra.  This representation
however
breaks when $\varphi$ is a multiple  of $2\gamma$ (recall $\gamma={\pi\over
x+1}$) or equivalently $e_\varphi$ is a multiple of $-2\alpha_0\alpha_-$.
 Setting
\eqn\reddef{\varphi=2n\gamma,\ n>S^z,}
${\cal H}_{2L}^{S^z}$ has an irreducible component $R_{2L}^{S^z,n}$
of dimension ${L-S^z\choose 2L}-{L-n\choose 2L}$.
The generating function \limdef\ reduced to this representation
reads
\eqn\limittwo{\hbox{Lim}\ tr_{R^{S^z,n}_{2L}}\exp(-T_I
\hat{\hbox{H}}_\varphi-iT_R
\hat{\hbox{P}}_\varphi)={\cal F}_n^{S^z}-{\cal F}_{S^z}^n\equiv K_n^{S^z},}
where recall that ${\cal F}_n^{S^z}$ refers to the right hand side of
\limitone. Eq. \limittwo\
corresponds to the trace of
$$
p^{L_0-c/24}\bar{p}^{\bar{L}_0-c/24},
$$
over the $\hbox{Vir}\otimes\bar{\hbox{Vir}}$  irreducible component of \virrep.
The matrix elements of the lattice generators \lattvir\ and \lattvirbar\
 between
states in the irreducible component $R_{2L}^{S^z,n}$  and states out of it
 vanish.
As an example,  the matrix elements of the lattice
Virasoro generators vanish exactly between the ground state $\hbox{u}$ (that
belongs to the irreducible component) and the
excited states $\hbox{v}$ and $\hbox{w}_1$  when $S^z=0$ and
$e_\varphi=-2\alpha_0\alpha_-$ (that do not belong to it). This is an exact
lattice analog of the
cancellations
in the action of Virasoro generators deduced from
 \stcft\ in $F_{\alpha,\alpha_0}$.

In the case when $x$ is rational , the representation theory of the periodic
Temperley-Lieb algebra becomes more involved, and the space $R_{2L}^{S^z,n}$
 becomes itself
reducible. Restrict for simplicity to the case $x+1$ integer. A new
irreducible component $\rho_{2L}^{S^z,n}$  appears then  of dimension
$$
\sum_{l\in Z}{L-S^z+l(x+1)\choose 2L}-{L-n+l(x+1)\choose 2L},
$$
where the sum truncates for negative arguments in the binomial
coeffcients. The generating function \limdef\ reduced to this representation
reads
\eqn\limitthree{\hbox{Lim}\ tr_{
\rho^{S^z,n}_{2L}}\exp(-T_I\hat{\hbox{H}}_\varphi-iT_R
\hat{\hbox{P}}_\varphi)=\sum_{r=1}^{x-1}\chi_{r,n-S^z}\bar{\chi}_{r,n+S^z},}
and corresponds of course to the trace of
$$
p^{L_0-c/24}\bar{p}^{\bar{L}_0-c/24},
$$
over the irreducible component of \virrep\ that is
%
\eqn\spirr{\bigoplus_{r=1}^{x-1}\hbox{Vir}_{r,n-S^z}\bar{\hbox{Vir}}_{r,n+S^z},}%
where in \limitthree\ $\chi_{rs}$ denotes the character of the Virasoro algebra
in the irreducible representation $\hbox{Vir}_{rs}$ and labels $r,s$ are
Kac labels parametrizing the conformal weights
$$
h_{rs}={[(x+1)r-xs]^2-1\over 4x(x+1)},
$$
see \ref\RC{A.Rocha-Caridi in ``Vertex operators in mathematics and physics",
eds. J.Lepowski, S.Mandelstam and I.Singer, Springer, New York (1985).} and
references therein.

Matrix elements of Temperley-Lieb generators, hence in particular of the
lattice Virasoro generators, vanish then exactly between
states in $\rho_{2L}^{S^z,n}$ and states out of it, in agreement with
corresponding results for Virasoro generators in $F_{\alpha,\alpha_0}$.
To analyze
further the irreducible representations in the rational case  it is
better to turn to the RSOS representation.

\newsec{Scaling, conjecture and numerical checks for the RSOS representation}

\subsec{Generalities}

We restrict for simplicity to $x$ integer. We recall the RSOS representation
of the
Temperley-Lieb algebra. The space ${\cal H}^{\rm RSOS}_{2L}$
is now spanned by vectors $|l_1,\ldots,l_{2L}>$
where the $l_i$ take values $l_i=1,\ldots,x$ with the constraint
 $|l_i-l_{i+1}|=1$. This space has dimension $\hbox{Tr}c^{2L}$ where
$c$ is the incidence matrix of the $A_{x}$ Dynkin diagram. One has
\eqn\tldefrsos{\eqalign{\left(e_j\right)_{l,l'}=&
\delta(l_1,l'_1)\ldots\delta(l_{j-1},
l'_{j-1})\delta(l_{j-1},l_{j+1})\delta(l_{j+1},l'_{j+1})\ldots\delta(l_{2L},
l'_{2L})\cr
&{\left(\sin(\pi l_j/(x+1))\sin(\pi l'_j/(x+1)\right)^{1/2}\over
\sin(\pi l_{j-1}/
(x+1))}\cr},}
with $l_{2L+1}\equiv l_1$. Define
\eqn\newhammi{\hat{\hbox{H}}=-{\gamma\over\pi\sin\gamma}\sum_{j=1}^{2L}
(e_j-e_\infty),}
and
\eqn\newpdef{\hat{\hbox{P}}={1\over i}\left({\gamma\over\pi\sin\gamma}\right)^2
\sum_{j=1}^{2L}[e_j,e_{j+1}].}
The space ${\cal H}^{\rm RSOS}$ decomposes as
\eqn\dechilbnew{{\cal H}^{\rm RSOS}=\bigoplus_{s=1}^{x}\rho^{0,s}_{2L},}
where the $\rho^{S^z,n}_{2L}$ are the irreducible representations of
the periodic
Temperley-Lieb algebra. One has from subsection 5.7
\eqn\limitfour{\hbox{Lim}\ tr_{\rho^{0,s}_{2L}}\exp(-T_I\hat{\hbox{H}}-iT_R
\hat{\hbox{P}})=\sum_{r=1}^{x-1}\chi_{rs}\bar{\chi}_{rs}.}
With the natural scalar product in ${\cal H}^{\rm RSOS}_{2L}$ one has now
$e_j^+=e_j$.

\subsec{Scaling limit}

It is well known that the eigenstates of \newhammi\ are a subset of the
 eigenstates
of \hammi\ for appropriate choices of the spin $S^z$ and the twist angle
 $\varphi$  \K \ref\BR{V.V.Bazhanov, N.Yu Reshetikhin, J.Mod.Phys. A4 (1989)
115.}
The same scaling limit can therefore be defined as in the vertex case.

\subsec{Conjecture}

We now state the main conjecture of this section. Introduce the same
 lattice quantities as before
\eqn\lattvirrsos{l_n={L\over 2\pi}\left\{-{\gamma\over\pi\sin\gamma}
\sum_{j=1}^{2L}
e^{inj\pi/L}\left(e_j-e_\infty
+{i\gamma\over\pi\sin\gamma}[e_j,e_{j+1}]\right)\right\}+{c\over 24}
\delta_{n,0},}
and
\eqn\lattvirbarrsos{\bar{l}_{-n}={L\over
2\pi}\left\{-{\gamma\over\pi\sin\gamma}
\sum_{j=1}^{2L}
e^{inj\pi/L}\left(e_j-e_\infty-{i\gamma\over\pi\sin\gamma}
[e_j,e_{j+1}]\right)
\right\}+{c\over 24}\delta_{n,0}.}
Consider the following double limit process. Choose a value of $s$
 and a pair of
integers $N,\bar{N}$.
Using the characterization of Bethe eigenstates by integers, or
simply by ordering the
eigenenergies, select for every $L$ the set of
gaps which, after multiplication by $T_R$
reproduce in the limit $L\rightarrow\infty$ the conformal weights
 $h_{\alpha,\alpha_0}+n$
and $\bar{h}_{\alpha,\alpha_0}+\bar{n}$ with $n\leq N,\bar{n}\leq\bar{N}$
 (for $L$ large enough
but finite all these states will be present: we implicitely suppose we
 always are in such
 a situation). Consider the action of the generators $l_n$ (resp.
$\bar{l}_n$) restricted to these
states. We conjecture that it furnishes in the limit $L\rightarrow\infty$ a
representation
 of $\hbox{Vir}$
 (resp. $\bar{\hbox{Vir}}$) restricted to
$$
\bigoplus_{n=0}^N\bigoplus_{\bar{n}=0}^{\bar{N}}\bigoplus_{r=1}^{x-1}
\left(\chi_{rs}\right)_n\left(\bar{\chi}_{rs}\right)_{\bar{n}},
$$
that is  more precisely a representation of $P\hbox{Vir}P$ (resp.
$P\bar{\hbox{Vir}}P$ where $P$ is the projector on the
above subspace).  Moreover we conjecture that the  natural scalar product in
${\cal H}_{2L}^{\rm RSOS}$
coincides in
the $L\rightarrow\infty$ limit
with the hermitean form for which $l_n^{+}=l_{-n}$.

\subsec{Numerical checks}

Numerical checks are much of the same nature as the ones given for the
vertex case. We shall give just a few examples.

First consider the state at vanishing momentum whose gap reproduces in the
limit \limitfour\ the weights $h=2,\bar{h}=0$. Call it $|\hbox{w}_2>$. One has
thus
\eqn\rri{|\hbox{w}_2>\mapsto\sqrt{{2\over 1-24\alpha_0^2}}L_{-2}|\hbox{u}>,}
where by $|\hbox{u}>$ we denote the ground state of the RSOS model. By
Virasoro commutation relations one expects
\eqn\rrii{<\hbox{w}_2,l_{-2}\hbox{u}>\rightarrow\sqrt{{c\over 2}}.}
Results for this quantity are given in table 18.

\newsec{The case of fixed boundary conditions}

So far we have dealt with systems without boundaries. This is the most
favorable
for numerical checks. On the other hand we can  only select combinations
of left and right Virasoro representations of the type
\eqn\fixbi{\sum_{r=1}^{x-1}\hbox{Vir}_{rs}\bar{\hbox{Vir}}_{rs'},}
using the lattice symmetries. The irreducible representations of left or right
algebras have to be selected by hand after the hamiltonian has been
diagonalized,
which is not too satisfactory. This problem can be solved by turning to fixed
boundary conditions. Indeed, following  \ref\BS{M.Bauer,
H.Saleur, Nucl. Phys. B320 (1989) 591.}
consider the  RSOS model of section 6 restricted to the space
${\cal H}_{2L}^{a/bc}$ spanned as before by vectors
$|l_1,\ldots,l_{2L}>$ with
\eqn\fixbii{l_1=a,\ l_{2L-1}=b,\ l_{2L}=c.}
Then one has
\eqn\fixbiii{\hbox{Lim }tr_{{\cal H}_{2L}^{a/bc}}
\exp(-T\hat{\hbox{H}})=\chi_{da},}
where
\eqn\fixbiv{d=\hbox{inf }(b,c).}
To get all the posible parities of character labels one can also
consider the RSOS model with an odd number of heights where the same formula
holds.
We can therefore easily select in the lattice model a space whose scaling
states
will correspond to a single representation of the Virasoro algebra. The
complete Virasoro action can be conjectured based on the arguments of section 2
using the same lattice stress energy tensor and the results
of \ref\CI{J.Cardy, Nucl. Phys. B240 (1984) 514.}. One expects  that
\eqn\fixbv{l_n={2L\over
\pi}\left\{-{\gamma\over\pi\sin\gamma}\sum_{j=1}^{2L-1}(e_j-e_\infty)
\cos{nj\pi\over 2L}+\left({\gamma\over\pi\sin\gamma}\right)^2\sum_{j=1}^{2L-2}
[e_j,e_{j+1}]\sin{nj\pi\over 2L}\right\}+{c\over 24}\delta_{n,0},}
tend to  the Virasoro generators with the limit process already outlined in
sections
4 and 6. We have preformed detailed numerical checks of this conjecture
as before. The convergence is not as good due to the free boundary conditions
but the results are satisfactory. Examples are given in table 19 to be compared
with table 18.

\newsec{Conclusion}

The double limit process in the scaling  can probably be overcome
by considering operators at different ``times". Indeed consider for instance
\eqn\timei{l_n(T)\equiv \exp\left(\pi{T\over L}n\right)e^{-T\hat{\hbox{H}}}
l_ne^{T\hat{\hbox{H}}},}
for say  the RSOS model. Consider some scaling state $|\hbox{u}>$
and act on it with
\eqn\timeii{l_n(T_1)l_m(T_2),}
with  $T_1<T_2$ and consider the limit $T_1,T_2,L\rightarrow\infty$ with
$T_{1(2)}/L\rightarrow t_{1(2)}$.
We expect that acting on $|\hbox{u}>$   will  give  the following results. The
first term
$e^{T_2\hat{\rm{H}}}$ produces simply a factor
 $e^{\pi T_2/L(h_u+\bar{h}_u)}|\hbox{u}>$. As
before action of $l_m$ couples $|\hbox{u}>$ to many other states. Except for
the state that corresponds to $L_m|\hbox{u}>$ all the matrix elements
vanish for large $L$, but before their sum could give finite contribution after
insertion of $L_n$.  However
we now have the exponential factor $e^{(T_1-T_2)\hat{\rm{H}}}$.
For non-scaling states  the value of  $\hat{\hbox{H}}$
is finite and therefore in the limit $T_1,T_2\rightarrow\infty$ their
contribution
is damped out exponentially fast. For scaling states
the series with generic term $e^{-\pi T(h+\bar{h})/L}$ converges
and therefore limit can be taken term by term. So we expect that all the
unwanted terms  will disappear to give simply the state corresponding
to $L_nL_m|\hbox{u}>$. Hence the reformulation
\eqn\timeiii{\eqalign{&\hbox{Lim}_{T_1,T_2,L\rightarrow\infty}
\left(l_n(T_1)l_m(T_2)-l_m(T_1)l_n(T_2)\right)|\hbox{u}>=\cr
&\hbox{Lim}_{T,L\rightarrow\infty}\left((n-m)l_{n+m}(T)+
{c\over 12}(n^3-n)\delta_{n+m}\right)|\hbox{u}>\cr}.}

The approach we have used is simple minded but we think it gives  nice
results. We have  observed with remarkable precision the built-up of the
central term , the null vectors structure,
the metric properties, and we believe have indeed a way of extracting the
Virasoro algebra from the lattice model. Numerically, the integrable
lattice models provide probably the best regularization for quantum field
theories in 1+1 dimensions, and we are not aware of works using more direct
regularizations (eg. discretization of a free boson) that would give comparable
results. To make this more interesting analytically
one would need to reproduce our computations exactly  using the
 Bethe ansatz.
This seems difficult but maybe not impossible. If the program can be carried
out, it will provide a bridge between the Virasoro algebra and lattice
integrability. In mathematical terms, we have a conjectured homomorphism
between the universal enveloping algebra of the Virasoro algebra and the
Temperley-Lieb algebra when the number of generators goes to infinity, a very
intriguing algebraic result.

\bigskip
\noindent{\bf Acknowledgments}: we thank M.Bauer who collaborated with us at
the early
stage of this project. HS thanks V.Dotsenko,
D.Haldane and  N.Yu Reshetikhin for many useful discussions. HS
and WMK were supported  by the Packard Foundation, the National
Young Investigator program  (grant NSF-PHY-9357207)
 and DOE  (grant DE-FG03-84ER40168).

\vfill
\eject
\noindent {\bf Figure captions}
\bigskip
\noindent Figure 1: Conventions for labelling the sites and edges of the
square lattice and its dual
\smallskip
\noindent Figure 2: A Potts model on the square lattice with $K_x\neq K_y$
can be considered as discretization of an isotropic continuum medium by
rectangles.
We characterize the rectangles by the anisotropy angle $\theta$.
\smallskip
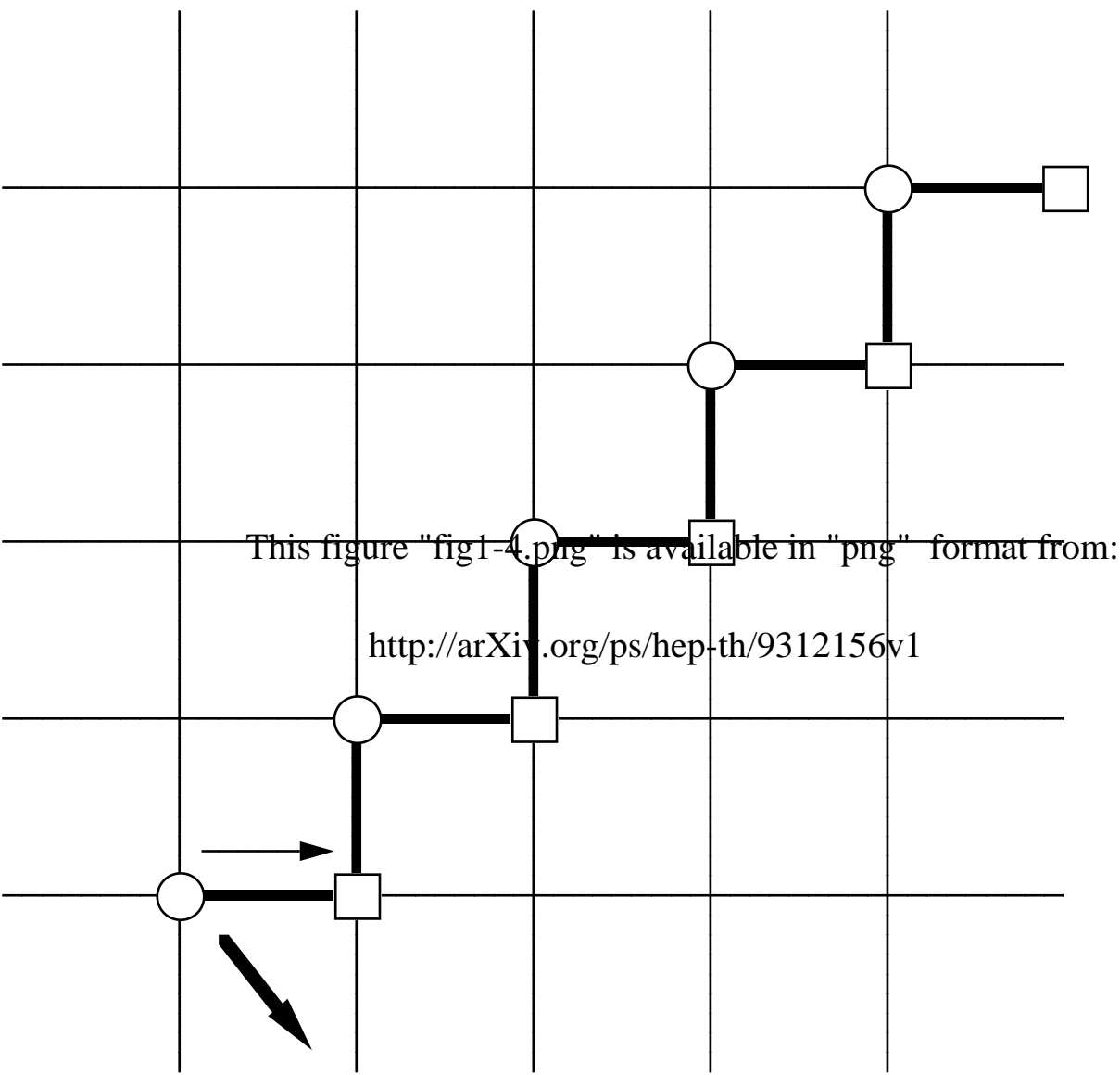
\noindent Figure 3: The operator $\hat{\tau}_D$ propagates in the direction of
the
light arrow. The operator ${\cal P}\hat{\tau}_D$ propagates in the direction of
the thick arrow.
\smallskip
\noindent Figure 4: When $K_x\neq K_y$,  ${\cal P}\hat{\tau}_D$ depends both
on $L_0+\bar{L}_0$ and $L_0-\bar{L}_0$.

\vfill
\eject

\hsize=29pc
\vsize=42pc

\vbox{\offinterlineskip
\def\space{height2pt&\omit&&\multispan9&}
\def\hline{\noalign{\hrule}}
\def\g{\hfil}
\halign{&\vrule#&\strut\quad #\hfil\quad\cr
\omit&\multispan{11} {\bf Table 1} \  Numerical values of
$-<\hbox{u},l_1 \hbox{v}>$
with $e_{\varphi}=0$ in the $S^z=0$ sector.$^a$\hfil&\omit\cr
\noalign{\vskip 4pt}
\hline
\space\cr
&\g$2L$\g&&\g $x=3$ \g&\omit&\g 4\g&\omit&\g 5\g&\omit&\g 6\g&\omit&\g 7\g&\cr
\hline
\space\cr
&\g \ 8\g&& 0.20266&\omit&0.15784&\omit&0.12955&\omit&0.11000&\omit&
0.094502&\cr
&\g 10\g&& 0.20330&\omit&0.15818&\omit&0.12974&\omit&0.11011&\omit&
0.095719&\cr
&\g 12\g&& 0.20361&\omit&0.15832&\omit&0.12979&\omit&0.11012&\omit&
0.095701&\cr
&\g 14\g&& 0.20379&\omit&0.15838&\omit&0.12979&\omit&0.11009&\omit&
0.095656&\cr
&\g 16\g&& 0.20390&\omit&0.15841&\omit&0.12977&\omit&0.11005&\omit&
0.095606&\cr
\space\cr
\hline
\space\cr
&\sevenrm \g extrapolation\g && 0.20416(2)&\omit&0.1584(3)&\omit&0.1292(2)
&\omit&0.1092(3)&\omit&0.0945(3)&\cr
\space\cr
\hline
\space\cr
&\sevenrm \g conjectured\g && 0.20412&\omit&0.15811&\omit&
0.12910&\omit&0.10911&\omit&0.094491&\cr
\space\cr
\hline
\noalign{\vskip 4pt}
\omit&\multispan{11} \sevenrm $^a$
Numerical values of $<\hbox{v},l_{-1}\hbox{u}>$ are identical.
 Conjectured numbers are $\sqrt{2}\alpha_0$\hfil&\omit\cr}}

\vskip 12pt           
\vbox{\offinterlineskip
\def\space{height2pt&\omit&&\multispan9&}
\def\hline{\noalign{\hrule}}
\def\g{\hfil}
\halign{&\vrule#&\strut\quad#\hfil\quad\cr
\omit&\multispan{11} {\bf Table 2} \  Numerical values of
$-<\hbox{u},l_1 \hbox{v}>$
with $e_{\varphi}=-2\alpha_o \alpha_-$ in the $S^z=0$ sector.$^b$\hfil&\omit\cr
\noalign{\vskip 4pt}
\hline
\space\cr
&\hfil$2L$\hfil&&\g $x=3$ \g&\omit&\g 4\g&\omit&\g 5\g&\omit&\g 6\g&\omit&
\g 7\g&\cr
\hline
height2pt&\omit&&\multispan9&\cr
&\g \ 8\g&& 0.40289&\omit&0.31422&\omit&0.25817&\omit&0.21938&\omit&
0.19088&\cr
&\g 10\g&& 0.40475&\omit&0.31521&\omit&0.25874&\omit&0.21973&\omit&
0.19109&\cr
&\g 12\g&& 0.40577&\omit&0.31570&\omit&0.25897&\omit&0.21982&\omit&
0.19111&\cr
&\g 14\g&& 0.40639&\omit&0.31597&\omit&0.25906&\omit&0.21982&\omit&
0.19106&\cr
&\g 16\g&& 0.40679&\omit&0.31612&\omit&0.25909&\omit&0.21978&\omit&
0.19098&\cr
\space\cr
\hline
\space\cr
&\sevenrm \g extrapolation\g&& 0.4082(2)&\omit&0.31638(5)&\omit&0.2591(2)&
\omit&0.2183(3)&\omit&0.1899(3)&\cr
\space\cr
\hline
\space\cr
&\sevenrm \g conjectured\g&& 0.40825&\omit&0.31623&\omit&0.25820&
\omit&0.21822&\omit&0.18898&\cr
\space\cr
\hline
\noalign{\vskip 4pt}
\omit&\multispan{11} \sevenrm $^b$ $<\hbox{v},l_{-1}\hbox{u}>$ is exactly zero
by TL representation theory.\hfil&\omit\cr}}

\vskip 12pt              
\vbox{\offinterlineskip
\def\space{height2pt&\omit&&\multispan3&&\multispan3&}
\def\hline{\noalign{\hrule}}
\def\g{\hfil}
\halign{&\vrule#&\strut\quad#\hfil\quad\cr
\omit&\multispan9 {\bf Table 3}\  Numerical values of
$<\hbox{v},l_{-1}\hbox{u}>$
in the $S^{z}=1$ sector.$^c$\hfil&\omit\cr
\noalign{\vskip 4pt}
\hline
\space\cr
&\omit&&\multispan3 \sevenrm \hfil $e_{\varphi}=0$\hfil&&\multispan3

\hfil $e_{\varphi}=2\alpha_o \alpha_-$ \hfil&\cr
\space\cr
&\omit&&\multispan7 \hrulefill&\cr
\space\cr
&\hfil$2L$\hfil&&\g $x=3$ \g&\omit&\g 7\g&&\g$x=3$\g&\omit&\g 7\g&\cr
\space\cr
\hline
\space\cr
&\g \ 8\g&& 0.37781&\omit&0.51645&&0.55468&\omit&0.60132&\cr
&\g 10\g&& 0.38728&\omit&0.52838&&0.57403&\omit&0.61729&\cr
&\g 12\g&& 0.39282&\omit&0.53742&&0.58494&\omit&0.62681&\cr
&\g 14\g&& 0.39634&\omit&0.54066&&0.59170&\omit&0.63299&\cr
&\g 16\g&& 0.39874&\omit&0.54416&&0.59619&\omit&0.63729&\cr
\space\cr
\hline
\space\cr
&\sevenrm \g extrapolation\g&& 0.4082(2)&\omit&0.547(2)&&0.6123(4)&
\omit&0.660(4)&\cr
\space\cr
\hline
\space\cr
&\sevenrm \g conjectured\g&& 0.40825&\omit&0.56695&&0.61237&
\omit&0.66144&\cr
\space\cr
\hline
\noalign{\vskip 4pt}
\omit& \multispan7 \sevenrm $^c$ The conjectured value is
$\sqrt{2}|\alpha_0-{e_\varphi\over 2}
\alpha_+ +{\alpha_-\over 2}|$.\hfil&\omit\cr}}

\vskip 12pt              
\vbox{\offinterlineskip
\def\space{height2pt&\omit&&\multispan3&&\multispan3&}
\def\hline{\noalign{\hrule}}
\def\g{\hfil}
\halign{&\vrule#&\strut\quad#\hfil\quad\cr
\omit&\multispan9 {\bf Table 4}\  Numerical values of
$-<\hbox{u},l_1 \hbox{v}>$ in the $S^{z}=1$ sector.$^d$\hfil&\omit\cr
\noalign{\vskip 4pt}
\hline
\space\cr
&\omit&&\multispan3 \sevenrm \hfil $e_{\varphi}=0$\hfil&&\multispan3 \hfil
 $e_{\varphi}=-2\alpha_o \alpha_-$ \hfil&\cr
\space\cr
&\omit&&\multispan7  \hrulefill &\cr  
\space\cr
&\hfil$2L$\hfil&&\g $x=3$ \g&\omit&\g 7\g&&\g$x=3$\g&\omit&\g 7\g&\cr
\space\cr
\hline
\space\cr
&\g \ 8\g&& 0.79920&\omit&0.72006&&0.97543&\omit&0.80498&\cr
&\g 10\g&& 0.80561&\omit&0.72945&&0.99188&\omit&0.81843&\cr
&\g 12\g&& 0.80922&\omit&0.73516&&1.00097&\omit&0.82630&\cr
&\g 14\g&& 0.81140&\omit&0.73901&&1.00646&\omit&0.83130&\cr
&\g 16\g&& 0.81281&\omit&0.74150&&1.01001&\omit&0.83469&\cr
\space\cr
\hline
\space\cr
&\sevenrm \g extrapolation\g&& 0.8169(3)&\omit&0.74632(1)&&1.021(1)&
\omit&0.849(2)&\cr
\space\cr
\hline
\space\cr
&\sevenrm \g conjectured\g&& 0.81650&\omit&0.75593&&1.02062&
\omit&0.85042&\cr
\space\cr
\hline
\noalign{\vskip 4pt}
\omit& \multispan9 \sevenrm $^d$ The conjectured value is
$\sqrt{2}|\alpha_0+{e_\varphi\over 2}
\alpha_+ -{\alpha_-\over 2}|$.\hfil&\omit\cr}}

\vskip 12pt              
\vbox{\offinterlineskip
\def\space{height2pt&\omit&&\multispan3&&\multispan3&}
\def\hline{\noalign{\hrule}}
\def\hsl{&\multispan9 \hrulefill &\cr}
\def\g{\hfil}
\halign{&\vrule#&\strut\quad#\hfil\quad\cr
\omit&\multispan{10} {\bf Table 5}\  Numerical values of $<w_{1},l_{-2}u>$
and $<w_{2},l_{-2}u>$ with $e_{\varphi}=0$.$^e$\hfil\cr
\noalign{\vskip 4pt}
\hsl
\space\cr
&\omit&&\multispan3 \sevenrm \hfil $<w_{2},l_{-2}u>$\hfil&&\multispan3
\hfil $<w_{1},l_{-2}u>$ \hfil&\cr
\space\cr
&\omit&\multispan8  \hrulefill &\cr  
\space\cr
&\hfil$2L$\hfil&&\g $x=3 \g$&\omit&\g 7\g&&\g$x=3$\g&\omit&\g 7\g&\cr
\space\cr
\hsl
\space\cr
&\g \ 8\g&& 0.70117&\omit&0.70264&&0.55799&\omit&0.26624&\cr
&\g 10\g&& 0.70357&\omit&0.70491&&0.56537&\omit&0.26839&\cr
&\g 12\g&& 0.70486&\omit&0.70612&&0.56934&\omit&0.26942&\cr
&\g 14\g&& 0.70562&\omit&0.70680&&0.57168&\omit&0.26992&\cr
&\g 16\g&& 0.70608&\omit&0.70721&&0.57318&\omit&0.27017&\cr
\space\cr
\hsl
\space\cr
&\sevenrm \g extrapolation\g&& 0.70705(2)&\omit&0.7073(3)&&0.57753(3)&
\omit&0.2704(2)&\cr
\space\cr
\hsl
\space\cr
&\sevenrm \g conjectured\g&& 0.70711&\omit&0.70711&&0.57735&
\omit&0.26726&\cr
\space\cr
\hsl
\noalign{\vskip4pt}
\omit&\multispan{10} \sevenrm $^e$ In the $S^{z}=0$ sector,
$<\hbox{u},l_2 \hbox{w}_2>=<\hbox{w}_{2},l_{-2} \hbox{u}>$, and $<\hbox{u},
l_{2}\hbox{w}_{1}>=-<\hbox{w}_{1},l_{-2}\hbox{u}>$.\hfil\cr
\noalign{\vskip 2pt}
\omit&\multispan{10} \sevenrm  Cojnectured  values are
respectively ${1\over\sqrt{2}}$ and $4\alpha_0$. \hfil\cr}}

\vskip 12pt              
\vbox{
\offinterlineskip
\def\space{height2pt&\omit&&\multispan3&&\multispan3&&\multispan3&}
\def\hline{\multispan{14}\hrulefill&\cr}
\def\g{\hfil}
\def\gp{\hskip 4pt}
\halign to \hsize {&\vrule# \gp 
&\gp \strut#\hfil\gp\cr
\omit&\multispan{13} {\bf Table 6}\  Numerical values of
$<\hbox{w}_{2},l_{-2}\hbox{u}>$,
$<\hbox{u},l_{2}\hbox{w}_{2}>$ and $-<\hbox{u},l_{2}\hbox{w}_{1}>$\hfil&\omit
\cr
\noalign{\vskip 2pt}
\omit&\multispan{13}{\hskip 42pt} with $e_{\varphi}=-2\alpha_0 \alpha_-$.$^f$
\hfil&\omit\cr
\noalign{\vskip 4pt}
\hline
\space\cr
&\omit&&\multispan3 \sevenrm \hfil $<\hbox{w}_{2},l_{-2}\hbox{u}>$\hfil&&
\multispan3 \hfil $<\hbox{u},l_{2}\hbox{w}_{2}>$\hfil&&
\multispan3 \hfil $-<\hbox{u},l_{2}\hbox{w}_{1}>$ \hfil&\cr
\space\cr
&\omit&\multispan{12}  \hrulefill &\cr  
\space\cr
&\hfil$2L$\hfil&&\g $x=3 \g$&\omit&\g 7\g&&\g$x=3$\g&\omit&\g 7\g&&\g$x=3$\g&
\omit&\g 7\g&\cr
\space\cr
\hline
\space\cr
&\g \ 8\g&& 0.76007&\omit&0.71725&&0.31235&\omit&0.60688&&
1.01588&\omit&0.51990&\cr
&\g 10\g&& 0.76243&\omit&0.71922&&0.31725&\omit&0.61099&&
1.03427&\omit&0.52490&\cr
&\g 12\g&& 0.76357&\omit&0.72022&&0.32000&\omit&0.61331&&
1.04453&\omit&0.52738&\cr
&\g 14\g&& 0.76415&\omit&0.72075&&0.32171&\omit&0.61475&&
1.05082&\omit&0.52869&\cr
&\g 16\g&& 0.76445&\omit&0.72104&&0.32285&\omit&0.61571&&
1.05495&\omit&0.52940&\cr
\space\cr
\hline
\space\cr
&\sevenrm \g extrapolation\g&& 0.76479(1)&\omit&0.72140(1)&&0.3274(6)&
\omit&0.6200(7)&&1.0692(3)&\omit&0.53029(2)&\cr
\space\cr
\hline
\space\cr
&\sevenrm \g conjectured\g&& 0.76376&\omit&0.71962&&0.32733&\omit&
0.62037&&1.06905&\omit&0.52523&\cr
\space\cr
\hline
\noalign{\vskip4pt}
\omit&\multispan{13} \sevenrm $^f$ In the $S^{z}=0$ sector, $<\hbox{w}_{1},
l_{-2}\hbox{u}>=0$
exactly. The conjectured results are respectively\hfil&\omit\cr
\noalign{\vskip 4pt}
\omit&\multispan{13} \sevenrm $\sqrt{{1+8\alpha_0^2\over 2}}$,
${1-24\alpha_0^2\over \sqrt{2+16\alpha_0^2}}$,
 ${8\alpha_0\over\sqrt{1+8\alpha_0^2}}$. \hfil&\omit\cr}}

\vskip 12pt              
\vbox{\offinterlineskip
\def\space{height2pt&\omit&&\multispan3&}
\def\hline{\noalign{\hrule}}
\def\s{\hskip 2cm}
\def\g{\hfil}
\halign{&\vrule#&\strut\quad#\hfil\quad\cr
\omit&\multispan5 {\bf Table 7}\  Numerical values
 of $<\hbox{u},l_{2}\hbox{w}_{1}><\hbox{w}_{1},l_{-2}\hbox{u}>+
<\hbox{u},l_{2}\hbox{w}_{2}><\hbox{w}_{2},l_{-2}\hbox{u}>$
 \hfil&\omit\cr
\noalign{\vskip 2pt}
\omit&\multispan5 \hskip 42pt with $e_{\varphi}=0$ and
$S^{z}=0$.$^g$\hfil&\omit\cr
\noalign{\vskip 6pt}
\hline
\space\cr
&\hfil$2L$\hfil&&\s \g $x=3 \g$&\omit&\s\g 7\g&\cr
\space\cr
\hline
\space\cr
&\g \ 8 \g&& \s 0.18029&\omit&\s 0.42282&\cr
&\g 10\g&& \s 0.17537&\omit&\s 0.42487&\cr
&\g 12\g&& \s 0.17269&\omit&\s 0.42601&\cr
&\g 14\g&& \s 0.17107&\omit&\s 0.42670&\cr
&\g 16\g&& \s 0.17002&\omit&\s 0.42715&\cr
\space\cr
\hline
\space\cr
&\sevenrm \g extrapolation \g&& \s 0.1667(3)&\omit&\s 0.4288(2)&\cr
\space\cr
\hline
\space\cr
&\sevenrm \g conjectured \g&& \s 0.16667&\omit&\s 0.42857&\cr
\space\cr
\hline
\noalign{\vskip 4pt}
\omit &\multispan5 \sevenrm $^g$ The conjectured value is
$1/2-16\alpha_0^2$.\hfil&\omit\cr}}
\vskip 12pt              
\vbox{\offinterlineskip
\def\space{height2pt&\omit&&\multispan9&}
\def\hline{\noalign{\hrule}}
\def\g{\hfil}
\halign{&\vrule#&\strut\quad#\hfil\quad\cr
\omit&\multispan{11} {\bf Table 8}\  Numerical values of
$<\hbox{u},l_{2}\hbox{w}_{1}><\hbox{w}_{1},l_{-2}\hbox{u}>+
<\hbox{u},l_{2}\hbox{w}_{2}><\hbox{w}_{2},l_{-2}\hbox{u}>$
 \hfil&\omit\cr
\noalign{\vskip 2pt}
\omit&\multispan{11} \hskip 42pt in the $S^{z}=0$ sector
 and $e_{\varphi}=-2\alpha_0\alpha_-$.$^{h}$ \hfil&\omit\cr
\noalign{\vskip 6pt}
\hline
\space\cr
&\hfil$2L$\hfil&&\g $x=3$ \g&\omit&\g 4\g&\omit&\g 5\g&\omit&\g 6\g&\omit&
\g 7\g&\cr
\space\cr
\hline
\space\cr
&\g \ 8 \g&& 0.23741&\omit&0.33642&\omit&0.38711&\omit&0.41659&
\omit&0.43529&\cr
&\g 10\g&&  0.24188&\omit&0.34123&\omit&0.39172&\omit&0.42096&
\omit&0.43944&\cr
&\g 12\g&&  0.24434&\omit&0.34388&\omit&0.39426&\omit&0.42336&
\omit&0.44172&\cr
&\g 14\g&&  0.24583&\omit&0.34550&\omit&0.39580&\omit&0.42480&
\omit&0.44308&\cr
&\g 16\g&&  0.24680&\omit&0.34655&\omit&0.39680&\omit&0.42573&
\omit&0.44395&\cr
\space\cr
\hline
\space\cr
&\sevenrm \g extrapolation \g&& 0.2501(3)&\omit&0.3501(3)&\omit&0.4002(3)&
\omit&0.4287(3)&\omit&0.4467(2)&\cr
\space\cr
\hline
\space\cr
&\sevenrm \g conjectured \g&& 0.25000&\omit&0.35000&\omit&0.40000&
\omit&0.42857&\omit&0.44643&\cr
\space\cr
\hline
\noalign{\vskip 4pt}
\omit&\multispan{11} \sevenrm $^h$ The conjectured value of $<\hbox{u},l_{2}
\hbox{w}_{1}><\hbox{w}_{1},l_{-2}\hbox{u}>+<\hbox{u},l_{2}\hbox{w}_{2}>
<\hbox{w}_{2},l_{-2}\hbox{u}>$ is equal to $c/2$.\hfil&\omit\cr}}

\vskip 12pt              
\vbox{\offinterlineskip
\def\space{height2pt&\omit&&\multispan3&&\multispan3&&\multispan3&}
\def\sp{height2pt&\omit&&\multispan7&&\multispan3&}
\def\hline{\multispan{14}\hrulefill&\cr}
\def\g{\hfil}
\def\gp{\hskip 4pt}
\halign to \hsize {&\vrule# \gp &\gp \strut#\hfil\gp\cr
\omit&\multispan{13} {\bf Table 9}\  Numerical values of
$<\hbox{u},l_1 \hbox{v}><\hbox{v},l_{1}\hbox{w}_{1}>$
and $<\hbox{u},l_1 \hbox{v}><\hbox{v},l_{1}\hbox{w}_{2}>$ .$^i$
\hfil&\omit\cr
\noalign{\vskip 4pt}
\hline
\sp\cr
&\omit&&\multispan7 \sevenrm \hfil $<\hbox{u},l_{1}\hbox{v}>
<\hbox{v},l_{1}\hbox{w}_{1}>$
 \hfil&&\multispan3
\hfil $<\hbox{u},l_{1}\hbox{v}><\hbox{v},l_{1}\hbox{w}_{2}>$\hfil&\cr
\sp\cr
&\omit&\multispan{12} \hrulefill &\cr
\space\cr
&\omit&&\multispan3 \sevenrm \hfil
 $e_{\varphi}=0$\hfil&&\multispan3 \hfil
 $e_{\varphi}=-2\alpha_o \alpha_-$
\hfil&&\multispan3 \hfil $e_{\varphi}=0$\hfil &\cr
\space\cr
&\omit&\multispan{12}  \hrulefill &\cr  
\space\cr
&\hfil$2L$\hfil&&\g $x=3$ \g&\omit&\g 7\g&&\g $x=3$\g&\omit&\g
7\g&&\g$x=3$\g&\omit&\g 7\g&\cr
\space\cr
\hline
\space\cr
&\g \ 8 \g&& -0.24115&\omit&-0.11157&&0.51308&\omit&0.22911&&
0.059246&\omit&0.013281&\cr
&\g 10\g&& -0.25788&\omit&-0.12095&&0.55112&\omit&0.24558&&
0.059256&\omit&0.013351&\cr
&\g 12\g&& -0.26721&\omit&-0.12535&&0.57256&\omit&0.25464&&
0.059247&\omit&0.013284&\cr
&\g 14\g&& -0.27291&\omit&-0.12798&&0.58576&\omit&0.26009&&
0.059229&\omit&0.013231&\cr
&\g 16\g&& -0.27663&\omit&-0.13188&&0.59444&\omit&0.26359&&
0.059208&\omit&0.013188&\cr
\space\cr
\hline
\space\cr
&\sevenrm \g extrapolation \g&& -0.289(1)&\omit&-0.1303(1)&&
0.625(3)&\omit&0.2714(2)&&0.05895(3)&\omit&0.01282(3)&\cr
\space\cr
\hline
\space\cr
&\sevenrm \g conjectured \g&& -0.28868&\omit&-0.13363&&0.62361&
\omit&0.27199&&0.058926&\omit&0.012627&\cr
\space\cr
\hline
\noalign{\vskip 4pt}
\omit&\multispan{13} \sevenrm $^i$ For $e_{\varphi}=0$, numerical calculation
 shows $<\hbox{w}_{1},l_{-1}\hbox{v}><\hbox{v},l_{-1}\hbox{u}>=-<\hbox{u},
l_{1}\hbox{v}><\hbox{v},l_{1}\hbox{w}_{1}>$ \hfil&\omit\cr
\noalign{\vskip 2pt}
\omit&\multispan{13} \sevenrm {\hskip 4pt} and $<\hbox{w}_{2},l_{-1}\hbox{v}>
<\hbox{v},l_{-1}\hbox{u}>=
<\hbox{u},l_{1}\hbox{v}><\hbox{v},l_{1}\hbox{w}_{2}>$. While
 for $e_{\varphi}=-2\alpha_{0}\alpha_{-}$, \hfil&\omit\cr
\noalign{\vskip 2pt}
\omit&\multispan{13}\sevenrm {\hskip 4pt}
$<\hbox{w}_{1(2)},l_{-1}\hbox{v}><\hbox{v},l_{-1}\hbox{u}>$ and
$<\hbox{u},l_{1}\hbox{v}>
<\hbox{v},l_{1}\hbox{w}_{2}>$ vanish  exactly. This is consistent \hfil&\omit
\cr
\noalign{\vskip 2pt}
\omit&\multispan{13} \sevenrm {\hskip 4pt} with $|\hbox{v}>\longrightarrow
L_{-1}|v_{\alpha,\alpha_0}>$ and $|\hbox{w}_{2}>\longrightarrow
 L_{-2}|v_{\alpha,\alpha_{0}}>$ respectively. The conjectured values
\hfil&\omit\cr
\noalign{\vskip 2pt}
\omit&\multispan{13} \sevenrm {\hskip 4pt}
are respectively $-2\alpha_0$, $4\alpha_0\sqrt{1+8\alpha_0^2}$,
$2\sqrt{2}\alpha_0^2$.\hfil&\omit\cr}}

\vskip 12pt       
\vbox{\offinterlineskip
\def\space{height2pt&\omit&&\multispan5&}
\def\hline{\noalign{\hrule}}
\def\s{\hskip 2cm}
\def\g{\hfil}
\halign{&\vrule#&\strut\quad#\hfil\quad\cr
\omit&\multispan7 {\bf Table 10} \  Numerical values of
 $<\hbox{w},\bar{l}_{-1} \hbox{v}><\hbox{v},l_{-1}\hbox{u}>$
with $e_{\varphi}=0$ and $S^{z}=0$.$^j$\hfil&\omit\cr
\noalign{\vskip 4pt}
\hline
\space\cr
&\hfil$2L$\hfil&&\omit&\omit&\s \g $x=3 \g$&\omit&\g\s 7\g&\cr
\hline
\space\cr
&\g \ 8 \g&&\omit&\omit&\hfil \s 0.040467&\omit&\s 8.9975E-3&\cr
&\g 10\g&&\omit&\omit&\hfil \s 0.040377&\omit&\s 8.9183E-3&\cr
&\g 12\g&&\omit&\omit&\hfil \s 0.040442&\omit&\s 8.8122E-3&\cr
&\g 14\g&&\omit&\omit&\hfil \s 0.040540&\omit&\s 8.7425E-3&\cr
&\g 16\g&&\omit&\omit&\hfil \s 0.040639&\omit&\s 8.6926E-3&\cr
\space\cr
\hline
\space\cr
&\sevenrm \g extrapolation \g&&\omit&\omit&\hfil \s 0.04157(2)&\omit&\s
8.713(7)E-3&\cr
\space\cr
\hline
\space\cr
&\sevenrm \g conjectured \g&&\omit&\omit&\hfil \s 0.041667&\omit&\s
8.9286E-3&\cr
\space\cr
\hline
\noalign{\vskip 4pt}
\omit&\multispan{7} \sevenrm $^j$ Conjectured value is $2\alpha_0^2$.
 \hfil&\omit\cr}}

\vskip 12pt       
\vbox{\offinterlineskip
\def\space{height2pt&\omit&&\multispan5&}
\def\hline{\noalign{\hrule}}
\def\s{\hskip 2cm}
\def\g{\hfil}
\halign{&\vrule#&\strut\quad#\hfil\quad\cr
\omit&\multispan{7} {\bf Table 11} \  Numerical values of
 $<\hbox{w}^{'},l_{-2}\hbox{u}^{'}>
-{3\over 2(2h_{21}+1)}<\hbox{w}^{'},l_{-1}\hbox{v}^{'}>
<\hbox{v}^{'},l_{-1}\hbox{u}^{'}>$\hfil&\omit\cr
\noalign{\vskip 2pt}
\omit&\multispan7 {\hskip 42pt} with
$e_{\varphi}=-2\alpha_0\alpha_-$.\hfil&\omit\cr
\noalign{\vskip 4pt}
\hline
\space\cr
&\hfil$2L$\hfil&&\omit&\omit&\s \g $x=3 \g$&\omit&\g\s 7\g&\cr
\hline
\space\cr
&\g \ 8 \g&&\omit&\omit&\hfil \s 0.25098&\omit&\s 0.21854&\cr
&\g 10\g&&\omit&\omit&\hfil \s 0.17640&\omit&\s 0.14104&\cr
&\g 12\g&&\omit&\omit&\hfil \s 0.12864&\omit&\s 0.09025&\cr
&\g 14\g&&\omit&\omit&\hfil \s 0.09722&\omit&\s 0.05808&\cr
&\g 16\g&&\omit&\omit&\hfil \s 0.07576&\omit&\s 0.03710&\cr
\space\cr
\hline
\space\cr
&\sevenrm \g extrapolation \g&&\omit&\omit&\hfil \s -0.00031(2)&\omit&\s
-0.00047(3)&\cr
\space\cr
\hline
\space\cr
&\sevenrm \g conjectured \g&&\omit&\omit& \s 0&\omit&\s 0&\cr
\space\cr}
\hrule}

\vskip 12pt              
\vbox{\offinterlineskip
\def\space{height2pt&\omit&&\multispan3&&\multispan3&}
\def\hline{\noalign{\hrule}}
\def\g{\hfil}
\halign{&\vrule#&\strut\quad#\hfil\quad\cr
\omit&\multispan9 {\bf Table 12}\  Numerical values of $\left(
<\hbox{u},l_3 \hbox{y}_1>^2+<\hbox{u},l_3 \hbox{y}_3>^2\right)^{1/2}$.
$^k$  \hfil&\omit\cr
\noalign{\vskip 6pt}
\hline
\space\cr
&\omit&&\multispan3 \sevenrm \hfil $e_{\varphi}=0$\hfil&&\multispan3 \hfil
$e_{\varphi}=-2\alpha_{0}\alpha_{-}$ \hfil&\cr
\space\cr
&\omit&&\multispan{7}  \hrulefill &\cr  
\space\cr
&\hfil$2L$\hfil&&\g $x=3$ \g&\omit&\g 7\g&&\g$x=3$\g\hfil&\omit&\g 7\g&\cr
\space\cr
\hline
\space\cr
&\g \ 8 \g&& 0.98806&\omit&0.48071&&1.71282&\omit&0.92484&\cr
&\g 10\g&& 1.01080&\omit&0.48505&&1.76984&\omit&0.93693&\cr
&\g 12\g&& 1.02535&\omit&0.48884&&1.80542&\omit&0.94619&\cr
&\g 14\g&& 1.03469&\omit&0.49135&&1.82827&\omit&0.95221&\cr
&\g 16\g&& 1.04092&\omit&0.49297&&1.84364&\omit&0.95614&\cr
\space\cr
\hline
\space\cr
&\sevenrm \g extrapolation \g&&
1.0512(5)&\omit&0.4958(1)&&1.852(8)&\omit&0.9634(1)&\cr
\space\cr
\hline
\space\cr
&\sevenrm \g conjectured \g&& 1.06066&\omit&0.49099&&1.89737&\omit&0.95669&\cr
\space\cr
\hline
\noalign{\vskip 4pt}
\omit&\multispan{9} \sevenrm $k$  Conjectured values are respectively
$3\sqrt{6}\alpha_0$ and ${6\sqrt{6}\alpha_0\over\sqrt{1+12\alpha_0^2}}$ for
$e_{\varphi}=0$ and $e_{\varphi}=-2\alpha_0\alpha_-$.\hfil&\omit\cr}}

\vskip 12pt              
\vbox{\offinterlineskip
\def\space{height2pt&\omit&&\multispan3&&\multispan3&&\multispan3&}
\def\sp{height2pt&\omit&&\multispan7&&\multispan3&}
\def\hline{\multispan{14}\hrulefill&\cr}
\def\g{\hfil}
\def\gp{\hskip 4pt}
\halign to \hsize {&\vrule#\gp &\gp\strut #\hfil\gp\cr
\omit&\multispan{13} {\bf Table 13}\  Numerical values of
$<\hbox{u},l_{3}\hbox{y}_{2}>$ and $<\hbox{y}_2,l_{-3}\hbox{u}>$
 .$^l$ \hfil&\omit\cr
\noalign{\vskip 4pt}
\hline
\sp\cr
&\omit&&\multispan7 \sevenrm \hfil $<\hbox{u},l_{3}\hbox{y}_{2}>$
\hfil&&\multispan3
\hfil $<\hbox{y}_2,l_{-3}\hbox{u}>$\hfil&\cr
\sp\cr
&\omit&\multispan{12} \hrulefill &\cr
\space\cr
&\omit&&\multispan3 \sevenrm \hfil $e_{\varphi}=0$\hfil&&\multispan3
\hfil $e_{\varphi}=-2\alpha_o \alpha_-$ \hfil&&\multispan3 \hfil
$e_{\varphi}=-2\alpha_0\alpha_-$\hfil &\cr
\space\cr
&\omit&\multispan{12}  \hrulefill &\cr  
\space\cr
&\hfil$2L$\hfil&&\g $x=3$ \g&\omit&\g 7\g&&\g$x=3$\g&\omit&\g 7\g&&\g$x=3$
\g&\omit&\g 7\g&\cr
\space\cr
\hline
\space\cr
&\g \ 8 \g&& 1.32379&\omit&1.32759&&0.53730&\omit&1.11815&&
1.50860&\omit&1.37583&\cr
&\g 10\g&& 1.35347&\omit&1.35656&&0.57124&\omit&1.11552&&
1.53200&\omit&1.41013&\cr
&\g 12\g&& 1.37152&\omit&1.37431&&0.58933&\omit&1.17635&&
1.54733&\omit&1.41744&\cr
&\g 14\g&& 1.38284&\omit&1.38543&&0.60025&\omit&1.18940&&
1.55707&\omit&1.42766&\cr
&\g 16\g&& 1.39030&\omit&1.39275&&0.60740&\omit&1.19803&&
1.56346&\omit&1.43439&\cr
\space\cr
\hline
\space\cr
&\sevenrm \g extrapolation \g&& 1.3986(7)&\omit&1.4016(4)&&0.6314(2)&
\omit&1.204(2)&&
1.57505(1)&\omit&1.457(1)&\cr
\space\cr
\hline
\space\cr
&\sevenrm \g conjectured \g&& 1.41421&\omit&1.41421&&0.63246&\omit&
1.23017&&
1.58114&\omit&1.45160&\cr
\space\cr
\hline
\noalign{\vskip 4pt}
\omit&\multispan{13} \sevenrm $^l$ Numerical calculation shows for
$e_{\varphi}=0$ $<\hbox{y}_{2},l_{-3}\hbox{u}>=<\hbox{u},l_{3}\hbox{y}_2>$.
Conjectured values for the \hfil&\omit\cr
\omit&\multispan{13} \sevenrm above are respectively $\sqrt{2}$,
${2-48\alpha_0^2\over \sqrt{2+24\alpha_0^2}}$ and
$\sqrt{2+24\alpha_0^2}$.
\hfil&\omit\cr}}

\vskip 12pt              
\vbox{\offinterlineskip
\def\space{height2pt&\omit&&\multispan9&}
\def\hline{\noalign{\hrule}}
\def\g{\hfil}
\halign{&\vrule#&\strut\quad#\hfil\quad\cr
\omit&\multispan{11} {\bf Table 14}\  Numerical values
of $<\hbox{u},l_{2}\hbox{w}_{1}><\hbox{w}_{1},l_{-2}\hbox{u}>+<\hbox{u}
,l_{2}\hbox{w}_{2}><\hbox{w}_{2},l_{-2}\hbox{u}>$
 \hfil&\omit\cr
\noalign{\vskip 2pt}
\omit&\multispan{11} \hskip 42pt computed using $h^{(2)}$,
 in the $S^{z}=0$ sector and $e_{\varphi}=-2\alpha_0\alpha_-$. \hfil&\omit\cr
\noalign{\vskip 6pt}
\hline
\space\cr
&\hfil$2L$\hfil&&\g $x=3$ \g&\omit&\g 4\g&\omit&\g 5\g&\omit&\g 6\g&\omit&
\g 7\g&\cr
\space\cr
\hline
\space\cr
&\g \ 8 \g&& 0.20264&\omit&0.28469&\omit&0.32633&\omit&0.35048&
\omit&0.36577&\cr
&\g 10\g&&  0.21879&\omit&0.30716&\omit&0.35190&\omit&0.37780&
\omit&0.39418&\cr
&\g 12\g&&  0.22797&\omit&0.31987&\omit&0.36630&\omit&0.39313&
\omit&0.41008&\cr
&\g 14\g&&  0.23366&\omit&0.32770&\omit&0.37514&\omit&0.40252&
\omit&0.41980&\cr
&\g 16\g&&  0.23741&\omit&0.33286&\omit&0.38094&\omit&0.40867&
\omit&0.42511&\cr
\space\cr
\hline
\space\cr
&\sevenrm \g extrapolation \g&& 0.250(1)&\omit&0.351(3)&\omit&0.401(2)&
\omit&0.429(2)&\omit&0.4332(1)&\cr
\space\cr
\hline
\space\cr
&\sevenrm \g conjectured \g&& 0.25000&\omit&0.35000&\omit&0.40000&\omit&
0.42857&\omit&0.44643&\cr
\space\cr
\hline}}

\vskip 12pt              
\vbox{\offinterlineskip
\def\space{height2pt&\omit&&\multispan9&}
\def\hline{\noalign{\hrule}}
\def\g{\hfil}
\halign{&\vrule#&\strut\quad#\hfil\quad\cr
\omit&\multispan{11} {\bf Table 15}\  Numerical values of
 $<\hbox{u},l_{2}\hbox{w}_{1}><\hbox{w}_{1},l_{-2}\hbox{u}>+<\hbox{u},l_{2}
\hbox{w}_{2}><\hbox{w}_{2},l_{-2}\hbox{u}>$
 \hfil&\omit\cr
\noalign{\vskip 2pt}
\omit&\multispan{11} \hskip 42pt computed using $h^{(3)}$ in the $S^{z}=0$
 sector and $e_{\varphi}=-2\alpha_0\alpha_-$. \hfil&\omit\cr
\noalign{\vskip 6pt}
\hline
\space\cr
&\hfil$2L$\hfil&&\g $x=3$ \g&\omit&\g 4\g&\omit&\g 5\g&\omit&\g 6\g&\omit&\g
7\g&\cr
\space\cr
\hline
\space\cr
&\g \ 8 \g&& 0.17297&\omit&0.24065&\omit&0.27457&\omit&0.29412&\omit
&0.30645&\cr
&\g 10\g&&  0.19789&\omit&0.27614&\omit&0.31542&\omit&0.33805&\omit
&0.35232&\cr
&\g 12\g&&  0.21270&\omit&0.29718&\omit&0.33959&\omit&0.36400&\omit
&0.37938&\cr
&\g 14\g&&  0.22209&\omit&0.31050&\omit&0.35486&\omit&0.38038&\omit
&0.39645&\cr
&\g 16\g&&  0.22837&\omit&0.31940&\omit&0.36507&\omit&0.39131&\omit
&0.40783&\cr
\space\cr
\hline
\space\cr
&\sevenrm \g extrapolation \g&& 0.252(2)&\omit&0.353(3)&\omit&0.403(4)&\omit
&0.432(4)&\omit&0.449(4)&\cr
\space\cr
\hline
\space\cr
&\sevenrm \g conjectured \g&& 0.25000&\omit&0.35000&\omit&0.40000&\omit
&0.42857&\omit&0.44643&\cr
\space\cr
\hline}}

\vskip 12pt     
\vbox{\offinterlineskip
\def\space{height2pt&\omit&&\multispan3&&\omit&}
\def\s{\hskip 20pt}
\def\hline{\noalign{\hrule}}
\def\g{\hfil}
\halign{&\vrule#&\strut\quad#\hfil\quad\cr
\omit&\multispan7 {\bf Table 16} \
 Comparision of $<l_{-1}\hbox{u},l_{-1}\hbox{u}>$
and $<l_{-1}\hbox{u},\hbox{v}><\hbox{v},l_{-1}\hbox{u}>$
with $x=3$ \hfil&\omit\cr
\noalign{\vskip 2pt}
\omit&\multispan7 {\hskip 42pt} and $S^{z}=0$.$^m$ \hfil&\omit\cr
\noalign{\vskip 4pt}
\hline
\space\cr
&\omit&&\multispan3 \sevenrm \hfil $e_{\varphi}=0$\hfil&&
\s $e_{\varphi}=-2\alpha_o \alpha_-$ \hfil&\cr
\space\cr
&\omit&&\multispan{5} \hrulefill &\cr
\space\cr
&\hfil $2L$\hfil&&\sevenrm
\hfil $<l_{-1}\hbox{u},l_{-1}\hbox{u}>$\hfil&\omit&\hfil$
<l_{-1}\hbox{u},\hbox{v}><\hbox{v},l_{-1}\hbox{u}>$\hfil&&
\hfil $<l_{-1}\hbox{u},l_{-1}\hbox{u}>$\hfil&\cr
\space\cr
\hline
\space\cr
&\g \ 8 \g&&0.041248&\omit&\s0.041069&&\s 6.0139E-5&\cr
&\g 10\g&&0.041579&\omit&\s0.041330&&\s 1.0574E-4&\cr
&\g 12\g&&0.041744&\omit&\s0.041459&&\s 1.3536E-4&\cr
&\g 14\g&&0.041833&\omit&\s0.041531&&\s 1.5382E-4&\cr
&\g 16\g&&0.041882&\omit&\s0.041575&&\s 1.6498E-4&\cr
\space\cr
\hline
\space\cr
&\sevenrm \g extrapolation \g&& 0.041946(2)&\omit&\s0.041656(2)&&\s
1.800(1)E-4&\cr
\space\cr
\hline
\space\cr
&\sevenrm \g conjectured \g&&\g ***\g&\omit&\s0.041667&&\g***\g&\cr
\space\cr
\hline
\noalign{\vskip 4pt}
\omit& \multispan7 \sevenrm $^m$ The numerical value of
$<l_{-1}\hbox{u},\hbox{v}>
<\hbox{v},l_{-1}\hbox{u}>$ is exactly zero for
$e_{\varphi}=-2\alpha_o \alpha_-$. For $e_\varphi=0$, \hfil&\omit\cr
\omit& \multispan7 \sevenrm the conjectured value is
$2\alpha_0^2$.\hfil&\omit\cr}}

\vskip 12pt     
\vbox{\offinterlineskip
\def\space{height2pt&\omit&&\multispan3&&\omit&}
\def\hline{\noalign{\hrule}}
\def\s{\hskip 20pt}
\def\g{\hfil}
\halign{&\vrule#&\strut\quad#\hfil\quad\cr
\omit&\multispan7 {\bf Table 17} \  Comparision of $<l_{-1}\hbox{u},
l_{-1}\hbox{u}>$
and $<l_{-1}\hbox{u},\hbox{v}><\hbox{v},l_{-1}\hbox{u}>$ with $x=7$
\hfil&\omit\cr
\noalign{\vskip 2pt}
\omit&\multispan7 {\hskip 42 pt} and  $S^{z}=0$.  \hfil&\omit\cr
\noalign{\vskip 4pt}
\hline
\space\cr
&\omit&&\multispan3 \sevenrm \hfil $e_{\varphi}=0$\hfil&&
\s $e_{\varphi}=-2\alpha_o \alpha_-$ \hfil&\cr
\space\cr
&\omit&&\multispan{5} \hrulefill &\cr
\space\cr
&\hfil $2L$ \hfil&&$<l_{-1}\hbox{u},l_{-1}\hbox{u}>$\hfil&\omit&
\hfil$<l_{-1}\hbox{u},\hbox{v}><\hbox{v},l_{-1}\hbox{u}>$\hfil&&
\hfil \s $<l_{-1}\hbox{u},l_{-1}\hbox{u}>$\hfil&\cr
\space\cr
\hline
\space\cr
&\g \ 8 \g&&9.1232E-3&\omit&\s9.1489E-3&&\s 2.8701E-5&\cr
&\g 10\g&&9.1174E-3&\omit&\s9.1621E-3&&\s 4.7919E-5&\cr
&\g 12\g&&9.1029E-3&\omit&\s9.1587E-3&&\s 5.8863E-5&\cr
&\g 14\g&&9.0883E-3&\omit&\s9.1502E-3&&\s 6.4679E-5&\cr
&\g 16\g&&9.0776E-3&\omit&\s9.1404E-3&&\s 6.7453E-5&\cr
\space\cr
\hline
\space\cr&\sevenrm \g extrapolation \g&&
9.0094(4)E-3&\omit&\s8.9645(5)E-3&&\s 6.64(2)E-5&\cr
\space\cr
\hline
\space\cr
&\sevenrm \g conjectured \g&&\g***\g&\omit&\s8.9286E-3&&\g***\g&\cr
\space\cr}
\hrule}

\vskip 12pt              
\vbox{\offinterlineskip
\def\space{height2pt&\omit&&\multispan9&}
\def\hline{\noalign{\hrule}}
\def\g{\hfil}
\halign{&\vrule#&\strut\quad#\hfil\quad\cr
\omit&\multispan{11} {\bf Table 18}\  Numerical values of $<\hbox{w}_{2},
l_{-2}\hbox{u}>$
computed using the RSOS representation.$^{n}$ \hfil&\omit\cr
\noalign{\vskip 6pt}
\hline
\space\cr
&\hfil$2L$\hfil&&\g $x=3$ \g&\omit&\g 4\g&\omit&\g 5\g&\omit&\g 6\g&\omit&
\g 7\g&\cr
\space\cr
\hline
\space\cr
&\g \ 8 \g&& 0.48725&\omit&0.58002&\omit&0.62218&\omit&0.64544&\omit&
0.65976&\cr
&\g 10\g&&  0.49182&\omit&0.58415&\omit&0.62587&\omit&0.64881&\omit&
0.66290&\cr
&\g 12\g&&  0.49431&\omit&0.58642&\omit&0.62790&\omit&0.65066&\omit
&0.66462&\cr
&\g 14\g&&  0.49581&\omit&0.58779&\omit&0.62913&\omit&0.65177&\omit&
0.66564&\cr
&\g 16\g&&  0.49679&\omit&0.58868&\omit&0.62992&\omit&0.65248&\omit&
0.66630&\cr
\space\cr
\hline
\space\cr
&\sevenrm \g extrapolation \g&&
0.5001(3)&\omit&0.5917(3)&\omit&0.6326(3)&\omit&
0.6547(2)&\omit&0.6683(2)&\cr
\space\cr
\hline
\space\cr
&\sevenrm \g conjectured \g&& 0.50000&\omit&0.59161&\omit&0.63246&\omit&
0.65465&\omit&0.66815&\cr
\space\cr
\hline
\noalign{\vskip 4pt}
\omit&\multispan{11} \sevenrm $^n$ In this representation
$<\hbox{u},l_2\hbox{w}_2>$
equals $<\hbox{w},l_{-2}\hbox{w}_2>$  and the conjectured value \hfil&\omit\cr
\noalign{\vskip 2pt}
\omit&\multispan{11} \sevenrm {\hskip 2pt}of
 $<\hbox{w}_{2},l_{-2}\hbox{u}>$ is equal to $\sqrt{c/2}$. \hfil&\omit\cr}}

\vskip 12pt       
\vbox{\offinterlineskip
\def\space{height2pt&\omit&&\multispan5&}
\def\hline{\noalign{\hrule}}
\def\s{\hskip 3cm}
\def\g{\hfil}
\halign{&\vrule#&\strut\quad#\hfil\quad\cr
\omit&\multispan{7} {\bf Table 19a} \  Numerical values of
 $<\hbox{w}_{2},l_{-2}\hbox{u}>$ computed in the RSOS representation
\hfil&\omit\cr
\noalign{\vskip 2pt}
\omit&\multispan7 {\hskip 42pt} with free boundary condition.$^o$\hfil&\omit\cr
\noalign{\vskip 4pt}
\hline
\space\cr
&\hfil$2L$\hfil&&\omit&\omit&\s \g $x=3 \g$&\omit&\s\g 4\g&\cr
\hline
\space\cr
&\g \ 8 \g&&\omit&\omit&\hfil \s 0.43912&\omit&\s 0.51191&\cr
&\g 10\g&&\omit&\omit&\hfil \s 0.45340&\omit&\s 0.52894&\cr
&\g 12\g&&\omit&\omit&\hfil \s 0.46289&\omit&\s 0.54052&\cr
&\g 14\g&&\omit&\omit&\hfil \s 0.46950&\omit&\s 0.54872&\cr
&\g 16\g&&\omit&\omit&\hfil \s 0.47429&\omit&\s 0.55476&\cr
\space\cr
\hline
\space\cr
&\sevenrm \g extrapolation \g&&\omit&\omit&\hfil \s 0.499(3)&\omit&\s
0.582(2)&\cr
\space\cr
\hline
\space\cr
&\sevenrm \g conjectured \g&&\omit&\omit&\hfil \s 0.50000&\omit&\s 0.59161&\cr
\space\cr
\hline
\noalign{\vskip 4pt}
\omit&\multispan{7} $^o$ The conjecture value is $\sqrt{c/2}$.\hfil&\omit\cr}}

\vskip 12pt       
\vbox{\offinterlineskip
\def\space{height2pt&\omit&&\multispan7&}
\def\hline{\noalign{\hrule}}
\def\s{\hskip 1cm}
\def\g{\hfil}
\halign{&\vrule#&\strut\quad#\hfil\quad\cr
\omit&\multispan{9} {\bf Table 19b} \  Numerical values of
 $<\hbox{w}_{2},l_{-2}\hbox{u}>$ computed in the RSOS \hfil&\omit\cr
\noalign{\vskip 2pt}
\omit&\multispan9 {\hskip 42pt} representation with free boundary
condition.\hfil&\omit\cr
\noalign{\vskip 4pt}
\hline
\space\cr
&\hfil$2L$\hfil&&\omit&\omit& \s\g $x=5 \g$&\omit&\s \g 6\g&\omit&\s \g 7\g&\cr
\hline
\space\cr
&\g \ 6 \g&&\omit&\omit& \s 0.51599&\omit&\s 0.53113&\omit&\s 0.54009&\cr
&\g \ 8\g&&\omit&\omit& \s 0.54219&\omit&\s 0.55789&\omit&\s 0.56714&\cr
&\g 10\g&&\omit&\omit& \s 0.56036&\omit&\s 0.57661&\omit&\s 0.58617&\cr
&\g 12\g&&\omit&\omit& \s 0.57283&\omit&\s 0.58953&\omit&\s 0.59933&\cr
&\g 14\g&&\omit&\omit& \s 0.59669&\omit&\s 0.61418&\omit&\s 0.62444&\cr
\space\cr
\hline
\space\cr
&\sevenrm \g extrapolation \g&&\omit&\omit& \s 0.60145(1)&\omit&\s
0.62022(7)&\omit&\s 0.6372(1)&\cr
\space\cr
\hline
\space\cr
&\sevenrm \g conjectured \g&&\omit&\omit&  \s 0.63246&\omit&\s 0.65465&\omit&
\s 0.66815&\cr
\space\cr}
\hrule}

\listrefs
\bye